\documentclass[a4paper,11pt]{article}
\pdfoutput=1 

\usepackage{jheppub} 

\usepackage[T1]{fontenc} 
\usepackage{subcaption}
\graphicspath{ {graphs/} }

\preprint{YITP-SB-2017-18}

\title{\boldmath The Mellin Formalism for Boundary CFT$_d$}

\author[]{Leonardo Rastelli,}
\author[]{Xinan Zhou}

\affiliation[]{C. N. Yang Institute for Theoretical Physics, Stony Brook University, Stony Brook, 11794, NY, USA}

\emailAdd{leonardo.rastelli@stonybrook.edu}
\emailAdd{xinan.zhou@stonybrook.edu}


\abstract{We extend the Mellin representation of conformal field theory (CFT) to allow for conformal boundaries and interfaces. 
We consider the simplest holographic setup dual to an interface CFT -- a brane filling an $AdS_{d}$ subspace of
$AdS_{d+1}$  --
and perform a systematic study of Witten diagrams in this setup.
 As a byproduct of our analysis,
 we show that geodesic Witten diagrams  in this geometry  reproduce  interface CFT$_d$ conformal blocks, generalizing the analogous statement for CFTs with no defects.
 }

\begin{document}
\maketitle
\flushbottom
\section{Introduction}

Conformal field theories (CFTs) are usually studied in position space, the most intuitive presentation for many practical purposes. Indeed the basic observables
of a CFT are the correlation functions of its local operators, which can be expressed as functions of conformally invariant cross ratios of the insertion points in $\mathbb{R}^d$.  
Naturally, most technical developments in CFT have been carried out in this language -- the calculation of conformal blocks (which are basic ingredients in the conformal bootstrap program) and the calculation of boundary correlators in AdS/CFT being just two indicative examples from a huge body of work.

A few years ago,  Mack \cite{Mack:2009mi} advocated to rewrite CFT correlators in Mellin space. By performing an integral Mellin transform, one trades the position space cross ratios for Mandelstam-like invariants. The Mellin transform of a $d$-dimensional
CFT correlator (henceforth, the ``Mellin amplitude'' ${\cal M}$)\footnote{We follow the terminology introduced by Mack, and define the Mellin amplitude ${\cal M}$ (note the calligraphic font) by stripping off  a certain product of Gamma functions from the mere Mellin transform, which is denoted by $M$.  See (\ref{mack}, \ref{Mackmellin}) below.}
shares several formal similarities with a $(d+1)$-dimensional  S-matrix element. For starters, the number of  kinematic invariants -- conformal cross ratios CFT$_d$
 and Mandelstam invariants in $\mathbb{R}^{d+1}$ -- is the same.  What's more, the analytic structure of ${\cal M}$ is reminiscent of the tree-level scattering amplitude in flat space. The Mellin amplitude has poles in correspondence with the exchanged operators, with factorization properties  controlled by the operator product expansion (OPE). 

The analytic properties of the Mellin amplitude are particularly simple  for large $N$ theories: ${\cal M}$ is a meromorphic function with poles in correspondence with just
the exchanged {\it single-trace} operators. If the large $N$ CFT$_d$  admits a gravity dual, Mellin amplitudes are naturally interpreted as  the ``scattering amplitudes'' for $AdS_{d+1}$.
This goes some way towards
explaining the formal similarity between Mellin amplitudes for CFT$_d$ correlators and S-matrix elements in one higher dimension. 
 Further validation of this picture is gained by considering the flat space limit of anti-de Sitter space.  Sending the  radius of $AdS_{d+1}$  to infinity,  and scaling operator dimensions appropriately, the Mellin amplitude yields the flat space scattering amplitude in  $\mathbb{R}^{d+1}$ \cite{Penedones:2010ue,Paulos:2016fap}\footnote{The precise limiting procedure is explained
 in the cited works. A distinction needs to be made according to whether one is interested in massless or massive scattering in flat space.  In the former case, the flat space
 S-matrix is obtained by taking a certain integral transform of the Mellin amplitude, while in the latter case one only needs to strip off a simple kinematic prefactor.}.
The idea  that CFT correlators  are the anti-de Sitter analog of the flat-space S-matrix is of course at the core of the whole holographic correspondence, but is somewhat obscured
by the usual position space representation. The Mellin representation makes this analogy  transparent and facilitates the calculation of holographic correlators. To wit,
Witten diagrams take a very simple form in Mellin space, both at tree level \cite{Penedones:2010ue,Paulos:2011ie,Fitzpatrick:2011ia,Costa:2014kfa,Goncalves:2014rfa} and for loops \cite{Aharony:2016dwx}.  In fact, the analytic structure of ${\cal M}$  is so constraining that it has recently been possible to obtain the four-point functions of one-half BPS operators of arbitrary weight in $AdS_5 \times S^5$ by imposing solely the constraints of maximal supersymmetry, analyticity and crossing \cite{Rastelli:2016nze} -- a feat that would be extremely hard to replicate by the traditional position space methods.  While Mellin methods are ideally suited in the holographic context, they have many other uses \cite{Paulos:2012nu,Fitzpatrick:2012cg,Costa:2012cb,Alday:2014tsa,Goncalves:2014ffa,Paulos:2016fap,Alday:2016htq}.
 One of their most striking applications so far has been the calculation of anomalous dimensions in the $\varepsilon$-expansion \cite{Gopakumar:2016wkt,Gopakumar:2016cpb,Dey:2016mcs}, using a peculiar version of the bootstrap constraints.

In this work, we extend the Mellin representation to CFTs whose conformal symmetry is partially broken by a boundary or a codimension one defect  (an interface). 
In the boundary case, one considers the theory in the Euclidean $d$-dimensional half-space, {\it i.e.}, with one Cartesian coordinate restricted to be positive, $x_\perp~\equiv~x^d ~\geqslant~0$;
in the interface case, one considers the full $\mathbb{R}^d$, but with the subspace $x^d =0$ playing a distinguished role (one may even have two different CFTs for $x_d >0$ and $x_d <0$). 
For a given choice of bulk CFT, there are in general several consistent  ways to introduce a conformal boundary or interface.\footnote{For a given bulk CFT,  specified by the spectrum and three-point couplings of its local operators, the boundary (interface)   CFT is determined by the following additional data: the spectrum of boundary (interface) operators, the boundary  (interface) three-point couplings,   and the bulk-to-boundary (interface) couplings. All these conformal data are subject to a system of crossing  constraints. The cleanest set up is that of two-dimensional rational CFTs, where a complete classification of the consistent boundary theories was achieved, starting with the classic
work of Cardy \cite{Cardy:1989ir, Cardy:1991tv}.} 
By the folding trick \cite{Bachas:2001vj} a defect can always be viewed as a  boundary in the tensor product of the two CFTs that live on either side
of the interface, so  without loss of generality we could use the terminology of Boundary CFT (BCFT). 
However, because  the simplest holographic setup is dual to an Interface
CFT (ICFT),  most of the paper will be phrased  in the language of ICFT.    

There are compelling motivations for developing the Mellin technology for
conformal theories with interfaces and boundaries. First, these theories are very interesting in their own right. They have important  physical applications in statistical mechanics and condensed matter physics (see, {\it e.g.},  \cite{Cardy:1984bb}), 
 formal field theory (see, {\it e.g.}, \cite{Gaiotto:2008sa} for supersymmetric examples)
 worldsheet string theory, (where D-branes are defined by boundaries on the string world sheet) and holography \cite{Karch:2000gx,DeWolfe:2001pq}, to give only an unsystematic sampling of a large literature.
 Second, boundary and interface conformal field theories are  a useful theoretical arena to develop  the bootstrap program\footnote{{The modern bootstrap program was initiated in \cite{Rattazzi:2008pe}}. For pedagogical review see, {\it e.g.}, \cite{Rychkov:2016iqz,Simmons-Duffin:2016gjk}. } \cite{Liendo:2012hy,Gliozzi:2013ysa,Gliozzi:2015qsa,Billo:2016cpy,Gliozzi:2016cmg,Liendo:2016ymz}, especially
if one's  goal is to gain analytic insight. Indeed, the simplest non-trivial ICFT correlator is the two-point function with two ``bulk'' insertions ({\it i.e.}, two operators inserted at $x_\perp  \neq 0$). Being 
a function of a single cross ratio, it is more tractable than  the four-point function in an ordinary CFT, which has two cross ratios.

The generalization of Mack's definition to scalar correlators in ICFT is conceptually straightforward, but involves a few technical novelties.
We consider the most general scalar case, a correlation function with $n$  bulk insertions and $m$ boundary insertions. In Mellin space, such a correlator has the kinematic structure of a scattering amplitude in $\mathbb{R}^{d+1}$ in the presence of a fixed target of codimension one, with $n$ particles scattering off the target from the bulk and $m$ particles with momenta parallel to the target. 
There is some freedom in one's choice of the Gamma factor prefactor that should be stripped off to define the ``Mellin amplitude ${\cal M}$''. Our choice is motivated by the desire to make
holographic calculations as natural as possible, so we fix the prefactor by requiring that the contact Witten diagrams are just {\it constants} in Mellin space.

The main part of the paper deals with calculations of holographic correlators.  Our  holographic framework is the simplest version of the Karch-Randall  setup \cite{Karch:2000gx,Karch:2001cw}. The dual geometry is  taken to be $AdS_{d+1}$ with a 
preferred $AdS_d$ subspace.  In string theory, this geometry can be obtained by taking the near horizon limit of a stack of   $N$ ``color'' D-branes, intersecting a single ``flavor'' brane along the interface.
At large $N$, the backreaction of the flavor brane can be ignored. Schematically, 
the effective action is taken to be
\begin{equation}\label{karchrandall}
S =  \int_{AdS_{d+1}}  {\cal L}_{\rm bulk} [\Phi_i]  + \int_{AdS_{d}} \left( \, {\cal L}_{\rm interface} [\phi_I] + {\cal L}_{\rm bulk/interface} [\Phi_i, \phi_I] \, \right) \,.
\end{equation}
where $\Phi_i$ denotes the fields that live in the full space $AdS_{d+1}$, while $\phi_I$ denotes the additional fields living on the $AdS_{d}$ brane.  The holographic dictionary
associates to $\Phi_i$ the local operators ${\cal O}_i$ of the bulk\footnote{Holographic  boundary CFTs suffer from the terminological nightmare that ``bulk'' and ``boundary'' have 
twofold meanings.  To minimize confusion, we will mostly use ``bulk'' in the meaning of this sentence, {\it e.g.}, to refer to the CFT operators that lives in the full $\mathbb{R}^d$, to be contrasted
to the interface or boundary operators that live at $x_\perp =0$. }  CFT$_d$, and to $\phi_I$ the operators living on the $(d-1)$-dimensional interface at $x_\perp = 0$. 

We perform a systematic study of Witten diagrams  in this geometry, focussing on external scalar operators.  
One of the highlights of our analysis is the demonstration that
 geodesic Witten diagrams correspond to conformal blocks, generalizing the analogous statement \cite{Hijano:2015zsa} for  CFTs with no defects. 
 We also give general expressions for contact diagrams and for two-point exchange Witten diagrams,
  in any $d$ and for general conformal dimensions.\footnote{
After submitting v1 of this paper to the ArXiv, we learnt that several special cases had already been computed in \cite{Aharony:2003qf}. We have checked that in those cases our results are completely compatible with theirs. We are grateful to Andreas Karch for bringing   this reference to our attention.}
 By construction, contact Witten diagrams with no derivative vertices are constant in Mellin space,
 while  the Mellin amplitudes of exchange Witten diagrams are  holomorphic functions with simple poles.  We also derive 
   the spectral representations for these exchange Witten diagrams.

The remainder of the paper is  organized as follows. In Section \ref{mellin} we use the embedding formalism to set up the Mellin representation for general scalar correlators in BCFT. In Section \ref{Wgeo} we show that conformal blocks are dual to geodesic Witten diagrams in a simple setup where an $AdS_d$ ``brane'' inside $AdS_{d+1}$ partially breaks the isometry. In Section \ref{Wcontact} we compute the contact Witten diagrams in the $AdS_{d}\subset AdS_{d+1}$ geometry. We warm up with the simplest cases and gradually build up the techniques to compute more complicated integrals. In Section \ref{Wexchange} we deal with the exchange Witten diagrams. We provide both the ``truncation method'' which works for theories with special spectra as well as the spectral representation method which works for generic conformal dimensions.  Some concluding remarks are offered in Section \ref{conclusions}.

\section{Mellin formalism}\label{mellin}

In this section we introduce the basic definitions that generalize the Mellin representation of conformal correlators to CFTs with boundaries and interfaces. 

\subsection{Conformal covariance in embedding space}

We start by
deriving the general form of the correlation function of $n$ bulk scalar operators and $m$ interface scalar operators. (For definiteness, we will use
the language appropriate to the interface case, but all formulae will be valid for the boundary case with the obvious modifications).
 We will use the standard Euclidean coordinates $x^{\mu}=(x^1,\ldots,x^{d-1},x_\perp)$ and place the interface at $x_\perp=0$. The coordinates parallel to the interface will be denoted as $\vec{x}$. As is familiar, a 
 convenient way to make the conformal symmetry manifest is to
  to lift this space to an  ``embedding space'' of dimension $d+2$ and signature $(-,+,+,\ldots)$. In the embedding space, points are labelled by lightcone coordinates which we denote as $P^A=(P^+, P^-,P^1,\ldots,P^d)$. The physical space has only $d$ coordinates and is restricted to be on a projective null cone in the embedding space,
\begin{equation}
P^AP_A=0\;\;\;\;\;\;{\rm with}\;\;\;\;\;\; P^A\sim \lambda P^A\;.
\end{equation}
The physical space coordinates $x$ are  related to the embedding space by the map
\begin{equation}
x^\mu=\frac{P^\mu}{P^+}\;.
\end{equation} 
Using the scaling freedom we can fix $P^+$ to be 1, so that 
\begin{equation}
P^A=(1,\vec{x}^2+x_\perp^2,\vec{x},x_\perp)\;.
\end{equation}
The conformal group $SO(d+1,1)$ which acts non-linearly on $x^\mu$ is now realized linearly as the Lorentz group on the embedding coordinates $P^A$. Conformal invariants in the physical space can be conveniently constructed from the embedding space as $SO(d+1,1)$ invariants. 

In the presence of the interface, the conformal group is broken down to the subgroup $SO(d,1)$. In the embedding space language, this can be conveniently described by introducing a fixed vector $B^A$,
\begin{equation}
B^A=(0,0,\vec{0},1)\;.
\end{equation} 
Points on the boundary uplift to vectors $\widehat P_A =(1,\vec{x}^2,\vec{x},0)$ that are transverse to $B^A$, $\widehat P_A B^A = 0$. 
The residual conformal transformations $SO(d,1)$ correspond to the linear transformations of $P^A$ that keep $B^A$ fixed.

In this paper we shall focus on  scalar operators.  Scalar operators are assumed to transform homogeneously under rescaling in the embedding space, 
\begin{equation}
O_\Delta(\lambda P)=\lambda^{-\Delta}O_\Delta(P)\;,\;\;\;\;\;\widehat{O}_{\widehat{\Delta}}(\lambda \widehat{P})=\lambda^{-\widehat{\Delta}}\widehat{O}_{\widehat{\Delta}}(\widehat{P})\;. 
\end{equation}
Here $O_\Delta$ is a bulk operator and $\widehat{O}_{\widehat{\Delta}}$ an interface operator, with $\Delta$ and $\widehat{\Delta}$  their respective conformal dimensions. (Hatted quantities
will always be interface quantities).

The correlator of $n$ bulk and $m$ interface operators,
\begin{equation}
\mathcal{C}_{n,m}\equiv\langle O_{\Delta_1}(P_1)\ldots O_{\Delta_n}(P_n)\widehat{O}_{\widehat{\Delta}_1}(\widehat{P}_1)\ldots\widehat{O}_{\widehat{\Delta}_m}(\widehat{P}_m)\rangle\; ,
\end{equation}
should be invariant under the residual $SO(d,1)$ symmetry and have the correct scaling weights when we rescale the embedding coordinate of each operator. There are only a handful of $SO(d,1)$ invariant structures,
\begin{eqnarray} \label{invariants}
-2 P_i\cdot P_j & =& (x_i-x_j)^2 \equiv (\vec x_i - \vec x_j)^2 + (x_{\perp i} - x_{\perp j})^2  \\
-2 P_i\cdot \widehat{P}_I &= &  (\vec x_i - \vec x_J)^2 + (x_{\perp i})^2\\
-2 \widehat{P}_I\cdot \widehat{P}_J & = &   (\vec x_I - \vec x_J)^2\\
P_i\cdot B &= & x_{\perp i}  \, ,
\end{eqnarray}
where $i = 1, \dots n$ and $I = 1, \dots m$. The most general form of the scalar correlator is
\begin{equation}\label{BCFTcorr}
\mathcal{C}_{n,m}= \left(\prod_{i<j}(-2P_i\cdot P_j)^{-\delta^0_{ij}}\prod_{i,I}(-2P_i\cdot \widehat{P}_I)^{-\gamma^0_{iI}}\prod_{I<J}(-2\widehat{P}_I\cdot \widehat{P}_J)^{-\beta^0_{IJ}}\, ,\prod_{i}(P_i\cdot B)^{-\alpha^0_{i}}\right)f(\xi_r)
\end{equation}
where the exponents must obey
\begin{equation}
\begin{split}
{}&\sum_{j}\delta^0_{ij}+\sum_I\gamma^0_{iI}+\alpha^0_i=\Delta_i\;,\\
{}&\sum_i\gamma^0_{iI}+\sum_J\beta^0_{IJ}=\widehat{\Delta}_{I}\;
\end{split}
\end{equation}
in order to give the correct scaling weights, while $f$ is an arbitrary function that depends on the  cross ratios $\xi_r$, which are ratios of the invariants (\ref{invariants}) with zero scaling  weights.

Let us also recall that anti de Sitter space  admits a simple description using the embedding coordinates. Euclidean $AdS_{d+1}$  is just the hyperboloid defined by the equation
\begin{equation}
Z^2=-R^2,\;\;\;\;\;\;\;\;\; Z^0>0,\;\;\;\;\;\;\;\;\; Z\in \mathbb{R}^{1,d+1}\;.
\end{equation}
We will usually set $R=1$. The Poincar\'e coordinates of $AdS_{d+1}$ are related to the embedding coordinates as
\begin{equation}
Z^A=\frac{1}{z_0}(1,z_0^2+\vec{z}^2+z_\perp^2,\vec{z},z_\perp)\;.
\end{equation}
\subsection{Mellin formalism: review of  the bulk case}

A quick review of the Mellin formalism for the standard bulk CFT case is in order.
Using the embedding formalism, it is easy to see that the general $n$-point correlator of scalar primary operators must take the form 
\begin{equation}
\mathcal{C}_{n}= \prod_{i<j}(-2P_i\cdot P_j)^{-\delta^0_{ij}}f(\xi_r) \, .
\end{equation}
Here $\xi_r$ are the conformally invariant cross ratios, of the form $\frac{(P_i\cdot P_j)(P_k\cdot P_l)}{(P_i\cdot P_l)(P_k\cdot P_j)}$, 
and to achieve the correct scaling
the exponents $\delta^0_{ij} \equiv \delta^0_{ji} $ must obey  the constraints
\begin{equation}  \label{Mellinconstraints}
\sum_{j \neq i}\delta^0_{ij}=\Delta_i\;.
\end{equation}
The number of independent cross ratios in a $d$-dimensional spacetime is
\begin{eqnarray}
& n\geq d+1:  & \qquad   nd-\frac{1}{2}(d+1)(d+2)\, , \\
& n< d+1:   & \qquad \frac{1}{2}n(n-3)\,.
\end{eqnarray}
Indeed, the configuration space of $n$ points which has $nd$  parameters, to which we must subtract the dimension
of the conformal group $SO(d+1,1)$, which is $\frac{1}{2}(d+1)(d+2)$. This gives the counting in the first line, which is valid so long as there is no subgroup
of $SO(d+1,1)$ that fixes the points, which is the case for $n \geq d+1$. For  $n< d+1$ there is a non-trivial stability subgroup. We 
can use conformal transformations to send two of the $n$ points to the origin and the infinity. The remaining $n-2$ points will define a hyperplane so the stability group is the
  group $SO(d+2-n)$ of rotations perpendicular to this hyperplane. After adding back the dimension of the stability subgroup we get  the counting in the second line.

Mack \cite{Mack:2009mi} suggested that  instead of taking $\delta^0_{ij}$ to be fixed, we should view them as variables $\delta_{ij}$ satisfying the same constraints, 
\begin{equation}
\delta_{ij} = \delta_{ji}  \, ,\quad \sum_{j \neq i}\delta_{ij}=\Delta_i \, , 
\end{equation}
and write  the correlator as an integral transform with respect to these variables. More precisely, one defines the following (inverse) Mellin transform for the {\it connected}\footnote{The disconnected part is a sum of powers of $P_i \cdot P_j$ and
its Mellin transform is singular.} part of the correlator,
\begin{equation}\label{mack}
\mathcal{C}^{\rm conn}_{n}=\int [d\delta_{ij}]\,  M(\delta_{ij})\, (-2P_i\cdot P_j)^{-\delta_{ij}} \, ,
\end{equation}
with the integration contours parallel to the imaginary axis. 
The correlator $f(\xi_r)$ is captured by the function $M(\delta_{ij})$, which Mack called the \textit{reduced} Mellin amplitude. One can solve the constraints (\ref{Mellinconstraints}) by introducing fictitious ``momentum'' variables $p_i$
\begin{equation}
\delta_{ij}=p_i\cdot p_j\; ,
\end{equation}
obeying the ``momentum conservation'' and on-shell conditions
\begin{equation}
\sum_{i=1}^n p_i=0\, , \qquad
p_i^2=-\Delta_i\;.
\end{equation}
If we assume that the fictitious momenta live in a $D$-dimensional spacetime, a simple counting tells that the
  number of independent Mandelstam variables   $\delta_{ij}= p_i \cdot p_j$ is  $\frac{n(n-3)}{2}$ for  $D > n$, and $n (D-1)  - D(D+1)/2$ for $D \leq n$.
Happily, this agrees with the counting of  conformal cross ratios if we take $D = d+1$. The integration in (\ref{mack}) should be understood as being performed over a set of independent
Mandelstam variables.

For example, for the four-point function, one can solve the constraints in terms of  the usual Mandelstam variables $s= - (p_1 + p_2)^2$, $t= - (p_1 + p_4)^2$ and $u = - (p_1 + p_3)^2$,
subject to the constraint
\begin{equation}
s + t + u = \Delta_1 + \Delta_2 + \Delta_3 + \Delta_4 \,.
\end{equation}
The main point of  definition (\ref{mack}) is that  the analytic properties of $M(\delta_{ij})$  are dictated by the operator product expansion.
Indeed, consider the OPE
\begin{equation}
O_i (P_i) O_j (P_j) = \sum_k  c_{ij}^{\; k} \, \left(  (-2 P_i\cdot P_j)^{-\frac{\Delta_i + \Delta_j - \Delta_k}{2} } O_k (P_i)    \,  + {\rm  descendants} \right) \, ,
\end{equation}
where for simplicity ${O}_k$ is taken to be a scalar operator.  To reproduce the leading behavior as $P_{12} \to 0$, $M$ must have a pole at $\delta_{ij} =  \frac{\Delta_i + \Delta_j  - \Delta_k}{2}$, as can be seen by closing the $\delta_{ij}$ integration
contour to the left of the complex plane. More generally,
the location of the leading pole is controlled by the twist $\tau$ of the exchanged operator ($\tau \equiv \Delta- \ell$, the conformal dimension minus the spin). Conformal descendants contribute an infinite sequence of satellite poles,
so that all in all for any primary operator $O_k$ of twist $\tau_k$  that contributes to the $O_i O_j$ OPE
the reduced Mellin amplitude $M(\delta_{ij})$ has poles at
\begin{equation}  \label{Mellinpoles}
\delta_{ij} =  \frac{\Delta_i + \Delta_j  - \tau_k - 2n}{2} \, , \quad n = 0, 1, 2 \dots \, .
\end{equation}
Mack further defined \textit{Mellin amplitude} $\mathcal{M}(\delta_{ij})$ by stripping off a product of Gamma functions,
\begin{equation}\label{Mackmellin}
\mathcal{M}(\delta_{ij})=\frac{M(\delta_{ij})}{\prod_{i<j}\Gamma[\delta_{ij}]}\, .
\end{equation} 
This is a convenient definition because ${\cal M}$ has simpler factorization properties. In particular, for the four-point function, the $s$-channel OPE $(x_{12} \to 0)$ implies
that the Mellin amplitude ${\cal M} (s, t)$ has poles in $s$ with residues that are {\it polynomials} of $t$. These {\it Mack polynomials}
depend on the spin of the exchanged operator, in analogy with the familiar partial wave expansion of  a flat-space S-matrix.\footnote{The analogy is not perfect, because each operator contributes an infinite of satellite poles, and because
Mack polynomials are significantly more involved than the Gegenbauer polynomials that appear in the usual flat-space partial wave expansion.}

A remarkable theorem about the spectrum of CFTs in dimension $d >2$ was proven in \cite{Fitzpatrick:2012yx,Komargodski:2012ek}. For any two primary operators $O_1$ and $O_2$ of twists $\tau_1$ and $\tau_2$, and for each non-negative integer $k$, 
the CFT must contain an infinite family of  so-called ``double-twist'' operators with increasing spin $\ell$ and twist approaching $\tau_1 + \tau_2 + 2k$ as $\ell \to \infty$ \cite{Komargodski:2012ek,Fitzpatrick:2012yx}.
This  implies that  the  Mellin amplitude   has infinite sequences of poles   accumulating at these asymptotic values of the twist, so strictly speaking it is not a meromorphic function.\footnote{In two dimensions, there are no double-twist families,
but one encounters a different pathology: the existence of infinitely many operators of the same twist, because Virasoro generators have twist zero.}

As emphasized by Penedones \cite{Penedones:2010ue}, 
a key simplification in the analytic structure of the Mellin amplitudes occurs in large $N$ CFTs, where the double-twist operators  are recognized as  the  usual double-trace operators. Thanks to large $N$ factorization, spin $\ell$ conformal primaries of the schematic form 
$: O_1 \Box^n \partial^\ell O_2 :$, where $O_1$ and $O_2$ are single-trace operators,
 have twist  $\tau_1 + \tau_2 + 2n + O(1/N^2)$\footnote{For definiteness, we are using the large $N$ counting appropriate to a theory with matrix degrees of freedom, {\it e.g.}, a $U(N)$
gauge theory. In other kinds of large $N$ CFTs the leading correction would have a different power -- for example, $O(1/N^3)$ in the $A_N$ six-dimensional (2, 0) theory, and $O(1/N)$ in two-dimensional
symmetric product orbifolds.}  for any $\ell$.  Recall also that the Mellin amplitude 
 is defined in terms of the {\it connected} part of the $n$-point correlator, which is of order $O(1/N^{n-2})$ for unit-normalized single-trace operators. Consider for simplicity the connected four-point function, which is of order $O(1/N^2)$.
 A little thinking shows that 
 the contribution of intermediate double-trace operators arises precisely at  $O(1/N^2)$, 
 so that to this order we can use their uncorrected dimensions.\footnote{If the external dimensions are such that $\Delta_1 + \Delta_2 - \Delta_3 - \Delta_4$ is an even integer, the product of Gamma functions contains {\it double} poles in $s$ 
 This generates a logarithmic term singularity in position space, which  signals an $O(1/N^2)$ anomalous dimension for  certain double-trace
 operators exchanged in the $s$-channel.  (Analogous statements hold of course in the $t$ and $u$ channels). See \cite{longpaper} for a detailed discussion.}  Remarkably, the poles corresponding to the exchanged  double-trace operators are precisely captured by the product of Gamma functions $\prod_{i<j} \Gamma(\delta_{ij})$ that Mack stripped
 off to define the Mellin amplitude ${\cal M}$. All in all, we conclude  that in a large $N$ CFT,  ${\cal M}$  {\it is}   a meromorphic function,
and that to  leading non-trivial large $N$ order  its poles are controlled by the exchanged {\it single-trace} operators.

 If the large $N$ CFT further admits a weakly coupled gravity dual in $AdS_{d+1}$, we can directly interpret the Mellin amplitude as a ``scattering amplitude'' in $AdS_{d+1}$. Recall that
 to the leading non-trivial large $N$ order, a holographic correlator is captured by a sum of tree-level Witten diagrams, which are the position space Feynman diagrams for the holographic perturbative expansion.
 The internal propagators of a Witten diagram, usually referred to as bulk-to-bulk propagators, are Green's functions that have the fastest allowed decay at the boundary $\partial AdS_{d+1}$. 
The external legs  of the diagram are  bulk-to-boundary propagators, which enforce the analog of Dirichlet boundary conditions at  $\partial AdS_{d+1}$. The special treatment of external legs
is analogous to the LSZ reduction to obtain the flat-space S-matrix -- we are instructed to put external legs on ``on-shell''. For future use, we record here the expression for the propagators
of a scalar field of mass $m^2 = \Delta (d-\Delta)$ (where we have set the AdS radius $R \equiv 1$). The bulk-to-boundary propagator ${\cal G}^{\Delta}_{B\partial}$ takes the simple form \cite{Witten:1998qj}
\begin{equation}
{\cal G}^{\Delta}_{B\partial}(P,Z) =  C_{B \partial}^{\Delta, d} \, G^{\Delta}_{B\partial}(P,Z) \, , \quad G^{\Delta}_{B\partial}(P,Z)=\left(\frac{1}{-2P\cdot Z}\right)^{\Delta}  \ , 
\end{equation}
where the normalization constant is given by \cite{Freedman:1998tz}
\begin{equation}
C_{B \partial}^{\Delta, d} = \frac{\Gamma(\Delta)(2\Delta-d)}{\pi^{d/2}\Gamma(\Delta-d/2)}\;.
\end{equation}
The normalized bulk-to-bulk propagator  ${\cal G}^\Delta_{BB}$ is given by  \cite{Inami:1985wu,Fronsdal:1974ew,Burgess:1985zz,Burges:1985qq}
\begin{eqnarray}\label{blkblkprop}
\nonumber{\cal G}^\Delta_{BB}(Z,W) & = & C^{\Delta, d}_{BB} {G}^\Delta_{BB}(Z,W) \,,  \\
 \nonumber{G}^\Delta_{BB}(Z,W) & =   & \left(-\frac{-2Z\cdot W-2}{4}\right)^{-\Delta}{}_2F_1\left(\Delta,\Delta-\frac{d}{2}+\frac{1}{2};2\Delta-d+1;\left(\frac{4}{-2Z\cdot W-2}\right)\right)\,, \\
 C_{BB}^{\Delta, d}  & =  & \frac{\Gamma(\Delta)\Gamma(\Delta-\frac{d}{2}+\frac{1}{2})}{(4\pi)^{(d+1)/2}\Gamma(2\Delta-d+1)}\,.
\end{eqnarray}
(To avoid cluttering in many equations, in the rest of the paper we will work with the {\it unnormalized} propagators  $G^{\Delta}_{B\partial}$ and ${G}^\Delta_{BB}$).
While Witten diagrams have a rather involved expression in position space, they look very simple in Mellin space.  
The basic building blocks are the contact Witten diagrams associated to an $n$-point bulk interactions with no derivatives,
\begin{equation}
D_{\Delta_1\ldots\Delta_n}=\int dZ \prod_i G^{\Delta_i}_{B\partial}(P_i,Z)\, ,
\end{equation}
which  for $n \geq 4$ are very non-trivial functions of the cross ratios. But as shown by \cite{Penedones:2010ue}, the Mellin amplitude of such a contact diagram is just a constant! It is then easy to show
that contact diagram associated to derivative vertices with $2m$ derivatives are {\it polynomials} of degree $m$ in the Mandelstam variables. Similarly, 
exchange diagrams simplify significantly. The
 Mellin amplitude for an $s$-channel exchange Witten diagram  takes the form
 \cite{Penedones:2010ue,Costa:2012cb}
 \begin{equation}
{\cal M}_{\Delta, J} (s, t) =  \sum_{m=0}^\infty   \frac{Q_{J, m} (t)}{s - (\Delta - J) - 2m}  + P_{J-1} (s, t) \, ,
 \end{equation}
where $\Delta$ and $J$ are the dimension and spin of the exchanged field, $Q_{J, m} (t)$ are polynomials of degree $J$ in $t$ 
and $P_{J-1} (s, t)$ is a polynomial in $s$ and $t$ of degree $J-1$.

\subsection{Mellin formalism for Interface CFTs}

After these preliminaries, we are ready to define  the Mellin representation for interface CFTs. Recalling that a scalar correlator  $\mathcal{C}_{n,m}$ with $n$ bulk and $m$ interface insertions takes the
general form (\ref{BCFTcorr}), it is natural to write it in terms of the following integral transform, 
\begin{equation}
\begin{split}
\mathcal{C}_{n,m}= {}&\int \prod_{i<j}d\delta_{ij}(-2P_i\cdot P_j)^{-\delta_{ij}}\prod_{i,I}d\gamma_{iI}(-2P_i\cdot \widehat{P}_I)^{-\gamma_{iI}}\prod_{I<J}d\beta_{IJ}(-2\widehat{P}_I\cdot \widehat{P}_J)^{-\beta_{IJ}}\\
{}&\times \prod_{i}d\alpha_i(P_i\cdot B)^{-\alpha_{i}} M(\delta_{ij},\gamma_{iI},\beta_{IJ},\alpha_i)\; .
\end{split}
\end{equation}
The variables  $\delta_{ij},\;\gamma_{iI},\;\beta_{IJ},\;\alpha_i$ are
constrained to obey
\begin{equation}\label{constraints}
\begin{split}
{}&\sum_{j}\delta_{ij}+\sum_I\gamma_{iI}+\alpha_i=\Delta_i\;,\\
{}&\sum_i\gamma_{iI}+\sum_J\beta_{IJ}=\widehat{\Delta}_{I}\, .
\end{split}
\end{equation}
A simple counting tells that there are
\begin{equation}
\frac{n(n-1)}{2}+\frac{m(m-1)}{2}+nm-m  \end{equation}
independent such variables, in one-to-one correspondence with the independent conformal cross ratios so long as the spacetime dimension is high enough,
namely for $d > n+m$.

 By a natural generalization of the case with no interface, the constraints  (\ref{constraints})
 can be  solved in terms of some fictitious momenta. We assign to each bulk operator a $(d+1)$-dimensional momentum $p_i$, to  each  interface operator a  $d$-dimensional momentum $\widehat{p}_I$
 and to the interface itself a $(d+1)$-dimensional momentum $\mathcal{P}$. 
 The  momenta need  to be conserved and on-shell,
\begin{equation}\label{conserve}
\sum_{i}p_i+\sum_I\widehat{p}_I+\mathcal{P}=0 \, , \quad
p_i^2=-\Delta_i\;,\;\;\;\;\; \widehat{p}_I^2=-\widehat{\Delta}_I \, .
\end{equation}
Moreover, the momenta of the interface operators  must be orthogonal to  $\mathcal{P}$
\begin{equation}\label{ortho}
\widehat{p}_I\cdot \mathcal{P}=0\;.
\end{equation}
  Then we can write
     \begin{equation} \label{parametrization}
\delta_{ij}=p_i\cdot p_j\;, \quad
\gamma_{iI}=p_i\cdot \widehat{p}_I\;, \quad 
\beta_{IJ}=\widehat{p}_I\cdot \widehat{p}_J\;, \quad
\alpha_{i}=p_i\cdot \mathcal{P}\;.\\
\end{equation}
From (\ref{conserve}) and (\ref{ortho}), we can replace $\mathcal{P}^2$ by $-\sum_i\mathcal{P}\cdot p_i$ and $\mathcal{P}\cdot p_i$ by $-\sum_jp_i\cdot p_j-\sum_I p_i\cdot \widehat{p}_I$. For the remaining bilinears there are still $m$ equations relating $p_i\cdot \widehat{p}_I$ to $\widehat{p}_I\cdot \widehat{p}_J$: $\sum_i p_i\cdot \widehat{p}_J+\sum_I \widehat{p}_I\cdot \widehat{p}_J=0$. So the number of independent momentum bilinears is 
\begin{equation}
\frac{n(n-1)}{2}+\frac{m(m-1)}{2}+nm-m \,  \quad {\rm if} \; d > n+m \, ,
\end{equation}
in agreement with the number of independent conformal cross ratios. This is the appropriate counting for $d > n+m$. For $d \leq n+m$, {\it both} the counting of independent Mandelstam invariants (\ref{parametrization}) and the counting of  conformal cross-ratios
for a configuration of $n$ bulk and $m$ interface operators give instead\begin{equation}
n d + m (d-1) - \frac{1}{2}d(d+1)     \, \quad {\rm if} \; d \leq n+m \,.
\end{equation}
Clearly, the parametrization (\ref{parametrization}) corresponds to  the kinematic setup of a scattering process off a fixed target, 
with $n$ particles having arbitrary momenta $p_i$, $m$ particles having momenta $\hat p_I$ parallel to the target and ${\cal P}$  the momentum transfer in the direction perpendicular of the infinitely heavy target.

Following Mack's terminology, we call $M$ 
the {\it reduced} Mellin amplitude. In our case, we wish to define the {Mellin amplitude} ${\cal M}$ by 
\begin{equation}\label{defmellingeneral}
\mathcal{M}=\frac{M(\delta_{ij},\gamma_{iJ},\beta_{IJ},\alpha_i)}{\prod_{i<j}\Gamma(\delta_{ij})\prod_{i,J}\Gamma(\gamma_{iJ})\prod_{I<J}\Gamma(\beta_{IJ})\prod_i\Gamma(\alpha_i)}\, \cdot \frac{\Gamma(-\mathcal{P}^2)}{\Gamma(-\frac{\mathcal{P}^2}{2})}\;.
\end{equation}
The Gamma functions 
 in the denominator of the first fraction are the counterpart of  the Gamma function  (\ref{Mackmellin}), accounting for the expected contributions
in a holographic  interface  theory. To wit, the poles in $\Gamma(\delta_{ij})$ correspond to double-trace bulk operators of the form ${\cal O}_i \Box^n {\cal O}_j$, the poles in  $\Gamma(\gamma_{iJ})$ to
double-trace bulk-interface operators of the form ${\cal O}_i \Box^n \hat {\cal O}_J$, and the poles in $\Gamma(\beta_{IJ})$ to double-trace interface-interface operators
of the form $\hat {\cal O}_I \Box^n \hat {\cal O}_J$. The poles in $\Gamma(\alpha_i)$ correspond to interface operators of the form $\partial_\perp^n {\cal O}_i (\vec x, x_\perp = 0)$,
{\it i.e.}, to the restriction to the interface of a bulk operator and  its normal derivatives -- these operators are indeed present in the simple holographic setup for ICFT
that we consider below.\footnote{In a holographic BCFT setup, one would need to impose boundary conditions that would remove some of these operators, {\it i.e.}, Dirichlet boundary conditions would remove ${\cal O}_i (\vec x, x_\perp = 0)$ while Neumann boundary conditions would remove $\partial_\perp {\cal O}_i (\vec x, x_\perp = 0)$.}
 The factor of ${\Gamma(-\mathcal{P}^2)}/{\Gamma(-\frac{\mathcal{P}^2}{2})}$ has a different justification.  We have introduced  it to ensure  that the Mellin amplitude of a contact Witten diagram is just a constant, as we will show in Section \ref{Wcontact} below. Note that if there are only interface operators ($n=0$),  the constraints imply $\mathcal{P} \equiv 0$. The additional factor ${\Gamma(-\mathcal{P}^2)}/{\Gamma(-\frac{\mathcal{P}^2}{2})}$  becomes simply $2$ and our definition of ${\cal M}$ reduces to Mack's, up to an overall normalization.

Let us specialize the formalism to the important case of two bulk insertions $(n=2)$ and no interface insertion $(m=0)$. We have
\begin{equation}
\begin{split} 
\langle O_{\Delta_1}O_{\Delta_2} \rangle={}&\int d[\alpha,\delta](-2P_1\cdot P_2)^{-\delta_{12}}(P_1\cdot B)^{-\alpha_{1}}(P_2\cdot B)^{-\alpha_{2}}\\
{}&\times\Gamma(\delta_{12})\Gamma(\alpha_1)\Gamma(\alpha_2)\frac{\Gamma(-\frac{\mathcal{P}^2}{2})}{\Gamma(-\mathcal{P}^2)}\mathcal{M}(\alpha,\delta)\;.
\end{split}
\end{equation}
The variables $\delta_{12},\; \alpha_1,\; \alpha_2$ must obey 
\begin{equation}
\delta_{12}+\alpha_1=\Delta_1\;,\;\;\;\;\;\delta_{12}+\alpha_2=\Delta_2 \, .
\end{equation}
The constraints can be solved using the  parameterization
\begin{equation}
\delta_{12}=-p_1\cdot p_2\;,\;\;\;\;\alpha_{1}=-p_1\cdot \mathcal{P}\;,\;\;\;\;\alpha_{2}=-p_2\cdot \mathcal{P} \, ,
\end{equation}
with the constraints
\begin{equation}
p_1+p_2+\mathcal{P}=0\;,\;\;\;\;p_1^2=-\Delta_1,\;\;\;\;\;p_2^2=-\Delta_2\;.
\end{equation}
These constraints leave only one independent variable that is bilinear in the momenta, namely $\mathcal{P}^2$, 
 \begin{equation}
\begin{split}
p_1\cdot p_2={}&\frac{\mathcal{P}^2+\Delta_1+\Delta_2}{2}\;,\\
p_1\cdot \mathcal{P}={}&\frac{-\mathcal{P}^2+\Delta_1-\Delta_2}{2}\;,\\
p_2\cdot \mathcal{P}={}&\frac{-\mathcal{P}^2-\Delta_1+\Delta_2}{2}\;.\\
\end{split}
\end{equation}
It  turns out to be convenient to make a change of variable
\begin{equation}
\mathcal{P}^2=2\tau-\Delta_1-\Delta_2 \, .
\end{equation}
The Mellin representation becomes
\begin{equation}\label{defmellin}
\langle O_{\Delta_1}O_{\Delta_2} \rangle=\frac{1}{(2x_{1,\perp})^{\Delta_1}(2x_{2,\perp})^{\Delta_2}}\int_{-i\infty}^{i\infty} d\tau \left(\frac{\eta}{4}\right)^{-\tau}\frac{\Gamma(\tau)\Gamma(\Delta_1-\tau)\Gamma(\Delta_2-\tau)}{\Gamma(\frac{1+\Delta_1+\Delta_2}{2}-\tau)}\mathcal{M}(\tau)\; ,
\end{equation}
where $\eta$ is the standard conformal cross ratio,
\begin{equation}\label{defeta}
\eta=\frac{(x_1-x_2)^2}{x_{1,\perp} x_{2,\perp}} = \frac{(\vec{x}_1-\vec{x}_2)^2+(x_{1,\perp} -x_{2,\perp})^2}{x_{1,\perp} x_{2,\perp}} \,.
\end{equation}

\section{Conformal blocks as geodesic Witten diagrams}\label{Wgeo}

We now turn to an analysis of ICFT correlators in the  holographic setup of 
 (\ref{karchrandall}). A systematic analysis of Witten diagrams in this geometry is presented in the following two sections. In this section we focus on the simplest case, the two-point function
 of bulk scalar operators, which has well-known decompositions into conformal blocks in both a bulk and  an interface channel, and demonstrate the equivalence between such conformal blocks 
   and  {\it geodesic} Witten diagrams \cite{Hijano:2015zsa}. After  a brief review 
of the two conformal block decompositions of the ICFT two-point function (section \ref{sub1}), we define 
Witten diagrams and the geodesic Witten diagrams in our holographic setup (section \ref{sub2}).  We then provide two proofs of the equivalence: one using the conformal Casimir equation (section \ref{sub3})
 and another from explicit evaluation of the integrals (section \ref{sub4}).

\subsection{Conformal blocks in ICFT} \label{sub1}

The  conformal block expansion is a basic tool in conformal field theory. The canonical example is the conformal block decomposition of a four-point function in a CFT with no defects:
each block captures the contribution of a primary operator and all its conformal descendants in a given OPE limit. The presence of an interface introduces additional OPE channels.
The simplest non-trivial case is the two-point function of bulk operators, which admits two distinct OPE limits, known as the bulk and the interface channels.

Recall  that the two-point function of scalar bulk operators takes the general form 
\begin{equation}
\langle O_{\Delta_1}(P_1)O_{\Delta_2}(P_2)\rangle=\frac{f(\eta)}{(2 P_1\cdot B)^{\Delta_1}(2 P_2\cdot B)^{\Delta_2}}=\frac{f(\eta)}{(2 x_{1,\perp})^{\Delta_1}(2 x_{2,\perp})^{\Delta_2}} \, ,
\end{equation}
where $\eta$ is the conformal cross ratio,
\begin{equation}
\eta=\frac{-2P_1\cdot P_2}{(2 P_1\cdot B)(2 P_2\cdot B)}=\frac{(x_1-x_2)^2}{x_{1,\perp} x_{2,\perp}}\;.
\end{equation}
In the {\it bulk channel}, we use the bulk OPE to merge the two  operators into another bulk operator,
\begin{equation}\label{bulkOPE}
O_{\Delta_1}(x)O_{\Delta_2}(y)=\frac{\delta_{\Delta_1\Delta_2}}{(x-y)^{2\Delta_1}}+\sum_{k}\lambda_kC[x-y,\partial_y]O_{\Delta_k}(y)\,,
\end{equation}
where $k$ labels the conformal primary fields. Here $\lambda_k$ is the OPE coefficient and the differential operator $C[x-y,\partial_y]$ is completely determined by the conformal symmetry. The bulk OPE reduces two-point functions to one-point functions which are fully determined by kinematics up to an overall constant,
\begin{equation}
\langle O_{\Delta}(y) \rangle=\frac{a_O}{(2y_\perp)^\Delta}\;.
\end{equation}
Conformal symmetry implies that only scalar operators have a non-vanishing one-point function in the presence of the interface. The couplings $a_O$ are part of the dynamical data of the theory. The contribution from each $O_{\Delta_k}(y)$ can be resummed into the bulk-channel conformal blocks \cite{McAvity:1995zd},
\begin{equation}\label{bulkcb}
f_{\rm bulk}(\Delta_k,\eta)=\left(\frac{\eta}{4}\right)^{\frac{\Delta_k-\Delta_1-\Delta_2}{2}}{}_2F_1(\frac{\Delta_k+\Delta_1-\Delta_2}{2},\frac{\Delta_k+\Delta_2-\Delta_1}{2};\Delta_k-\frac{d}{2}+1;-\frac{\eta}{4})\;.
\end{equation}
Substituting (\ref{bulkOPE}) into the two-point function and using (\ref{bulkcb}) we arrive at the conformal block decomposition in the bulk channel,
\begin{equation}
f(\eta)=\left(\frac{\eta}{4}\right)^{-\Delta_1}\delta_{\Delta_1\Delta_2}+\sum_k\lambda_ka_kf_{\rm bulk}(\Delta_k,\eta)\;.
\end{equation}
Alternatively, (\ref{bulkcb}) can be obtained from solving the conformal Casimir equation \cite{Liendo:2012hy}
\begin{equation}
L_{\rm bulk}^2\langle O_{\Delta_1}(P_1)O_{\Delta_2}(P_2)\rangle=-C_{\Delta,0}^{\rm bulk}\langle O_{\Delta_1}(P_1)O_{\Delta_2}(P_2)\rangle
\end{equation}
with the boundary condition
\begin{equation}
f(\eta)\sim \eta^{\frac{\Delta-\Delta_1-\Delta_2}{2}}\;,\quad\quad \eta\to0\;.
\end{equation}
Here the eigenvalue $C^{\rm bulk}_{\Delta,\ell}=\Delta(\Delta-d)+\ell(\ell+d-2)$ where $\Delta$ and $\ell$ are respectively the conformal dimension and spin of the exchanged operator. The Casimir operator
\begin{equation}
L_{\rm bulk}^2=(L_1^{AB}+L_2^{AB})^2\equiv\frac{1}{2}(L_1^{AB}+L_2^{AB})(L_{1,AB}+L_{1,AB})
\end{equation}
is constructed from the $SO(d+1,1)$ generators
\begin{equation}
L_{AB}=P_A\frac{\partial}{\partial P^B}-P_B\frac{\partial}{\partial P^A}\;.
\end{equation}
In the {\it interface channel} we use the bulk-to-boundary OPE, writing a bulk operator as an infinite sum of interface operators
\begin{equation}
O_\Delta(x)=\frac{a_O}{(2x_\perp)^\Delta}+\sum_l \mu_l D[x_\perp,\partial_{\vec{x}}]\widehat{O}_{\Delta_l}(\vec{x})
\end{equation}
with $l$ labeling interface primary operators and $D[x_\perp,\partial_{\vec{x}}]$ a  differential operator completely fixed by symmetry. The residual conformal group $SO(d,1)$ also fixes the interface two-point function
\begin{equation}
\langle \widehat{O}_{\Delta_l}(\vec{x})\widehat{O}_{\Delta_l}(\vec{y})\rangle=\frac{1}{|\vec{x}-\vec{y}|^{2\Delta_l}}\;.
\end{equation}
The {interface-channel conformal block} is determined by resumming the contributions of the descendants of $\widehat{O}_{\Delta_l}$
and has the compact expression \cite{McAvity:1995zd} 
\begin{equation}\label{boundarycb}
f_{\rm interface}(\Delta_l,\eta)=\left(\frac{\eta}{4}\right)^{-\Delta_l}{}_2F_1(\Delta_l,\Delta_l-\frac{d}{2}+1;2\Delta_l+2-d;-\left(\frac{\eta}{4}\right)^{-1})\;.
\end{equation}
Using the interface conformal blocks (\ref{boundarycb}), the two-point function is decomposed as
\begin{equation}
f(\eta)=a_O^2+\sum_l\mu_l^2f_{\rm interface}(\Delta_l,\eta)\;.
\end{equation}
We can also derive (\ref{boundarycb}) using the Casimir equation \cite{Liendo:2012hy}. It must satisfy 
\begin{equation}
L_{\rm interface}^2\langle O_{\Delta_1}(P_1)O_{\Delta_2}(P_2)\rangle=-C_{\Delta,0}^{\rm interface}\langle O_{\Delta_1}(P_1)O_{\Delta_2}(P_2)\rangle\, ,
\end{equation}
with the boundary condition
\begin{equation}
f(\eta)\sim \eta^{-\Delta}\;,\quad\quad \eta\to\infty\;.
\end{equation}
Here $C_{\Delta,0}^{\rm interface}=\Delta(\Delta-d+1)+\ell(\ell+d-3)$. The Casimir operator $L_{\rm interface}^2$ is restricted to the unbroken conformal subgroup $SO(d,1)$
\begin{equation}
L_{\rm interface}^2=(\hat{L}_1^{\hat{A}\hat{B}})^2\equiv\frac{1}{2}\hat{L}_1^{\hat{A}\hat{B}}\hat{L}_{1,\hat{A}\hat{B}} \, ,
\end{equation}
where $\hat{L}_{\hat{A}\hat{B}}$ are the $SO(d,1)$ generators
\begin{equation}
\hat{L}_{\hat{A}\hat{B}}=P_{\hat{A}}\frac{\partial}{\partial P^{\hat{B}}}-P_{\hat{B}}\frac{\partial}{\partial P^{\hat{A}}}\;.
\end{equation}

\subsection{Witten diagrams and geodesic Witten diagrams}
We now turn to the holographic setup of (\ref{karchrandall}) and introduce the different kinds of Witten diagrams that we are going to study in the rest of the paper.
We consider only tree level diagrams.

\label{sub2}
\begin{figure}[t]
 
\begin{subfigure}{0.32\textwidth}
\includegraphics[width=0.9\linewidth]{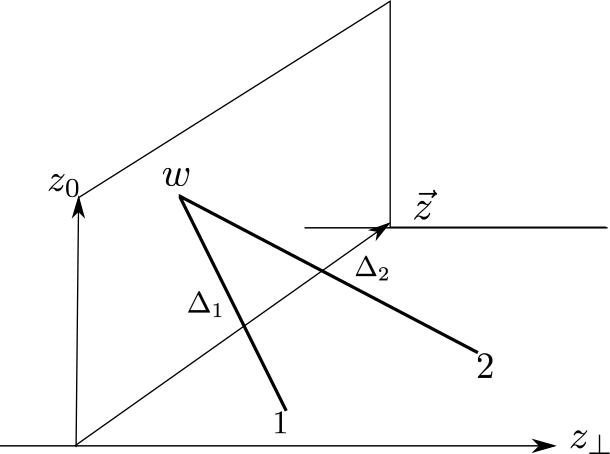} 
\caption{contact}
\label{fcontact}
\end{subfigure}
\begin{subfigure}{0.32\textwidth}
\includegraphics[width=0.9\linewidth]{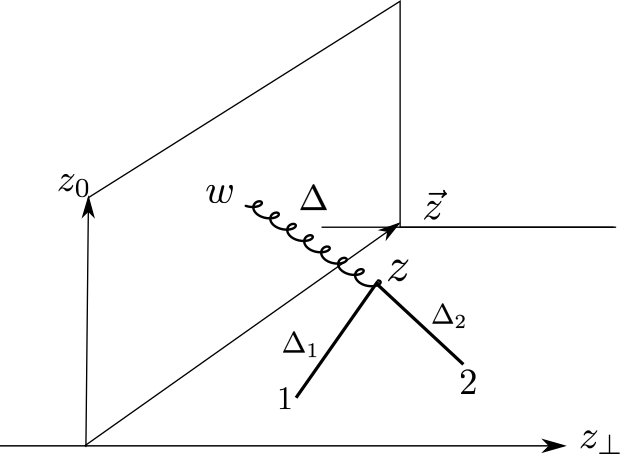}
\caption{bulk exchange}
\label{fbulkx}
\end{subfigure}
\begin{subfigure}{0.32\textwidth}
\includegraphics[width=0.9\linewidth]{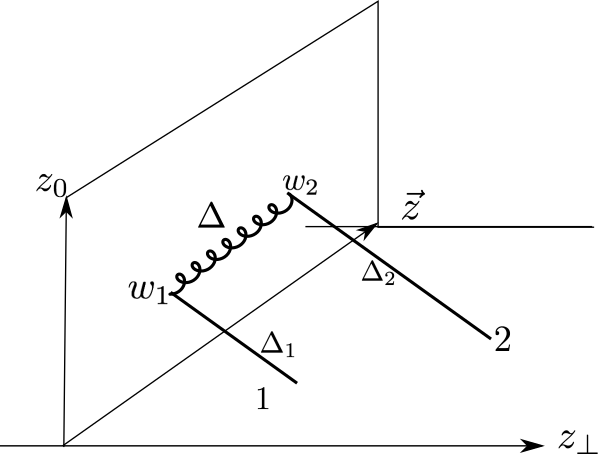}
\caption{interface exchange} 
\label{fboundaryx}
\end{subfigure}
 
\caption{Three types of Witten diagrams}
\label{fWitten}
\end{figure}

First of all, there are $(n+m)$-point {\it contact} Witten diagrams with  $n$ operators on the boundary of $AdS_{d+1}$ and $m$ operators on the boundary of $AdS_d$.  For example, Figure \ref{fcontact} represents the case $n=2$, $m=0$.  Such Witten diagrams involve vertices of the type $\Phi_{i_1}\ldots\Phi_{i_n}\phi_{I_1}\ldots \phi_{I_m}\subset \mathcal{L}_{\rm bulk/interface}$ in (\ref{karchrandall}). We denote them as
\begin{equation}
W_{n,m}(P_i,\hat{P}_I)=\int_{AdS_d}dW\prod_{i=1}^{n}\prod_{I=1}^{m} G_{B\partial}^{\Delta_i}(P_i,W)G_{B\partial}^{\hat{\Delta}_I}(\hat{P}_I,W)\;.
\end{equation}
We will also consider two-point exchange diagrams. A bulk exchange Witten diagram is drawn in Figure \ref{fbulkx} and we define it as
\begin{equation}\label{wdbulk}
W_{\rm bulk}(P_1,P_2)=\int_{AdS_{d+1}}dZ\int_{AdS_d}dW G_{B\partial}^{\Delta_1}(P_1,Z)G_{B\partial}^{\Delta_2}(P_2,Z)G_{BB}^{\Delta}(Z,W)\;.
\end{equation}
Such a Witten diagram involves the cubic vertex $\Phi_{i_1}\Phi_{i_2}\Phi_{i_3}\subset \mathcal{L}_{\rm bulk}$ and the tadpole vertex $\Phi_i\subset\mathcal{L}_{\rm bulk/interface}$.
An interface exchange Witten diagram is depicted  in Figure \ref{fboundaryx} and we define it as 
\begin{equation}\label{wdboundary}
W_{\rm interface}(P_1,P_2)=\int_{AdS_d}dW_1\int_{AdS_d}dW_2G_{B\partial}^{\Delta_1}(P_1,W_1)G_{BB}^{\Delta}(W_1,W_2)G_{B\partial}^{\Delta_2}(P_2,W_2)\;.
\end{equation}
Such a Witten diagram involves vertices of the type $\Phi_i\phi_I\subset\mathcal{L}_{\rm bulk/interface}$ and $\phi_{I_1}\phi_{I_2}\subset\mathcal{L}_{\rm interface}$.

\begin{figure}[t]
\begin{center}
\includegraphics[scale=0.6]{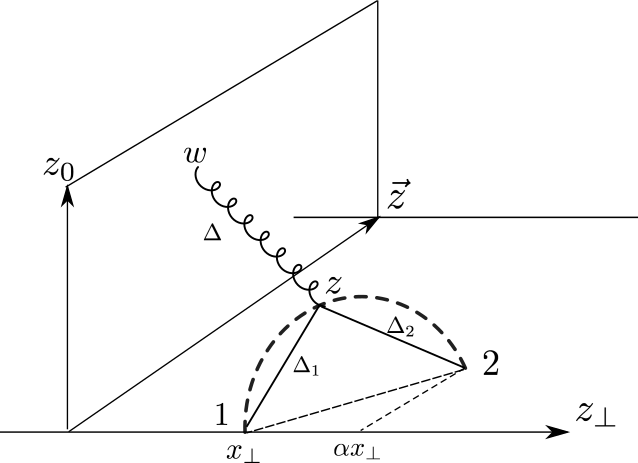}
\caption{The bulk-channel geodesic Witten diagram.}
\label{fbulkgeo}
\end{center}
\end{figure}

\begin{figure}[t]
\begin{center}
\includegraphics[scale=0.6]{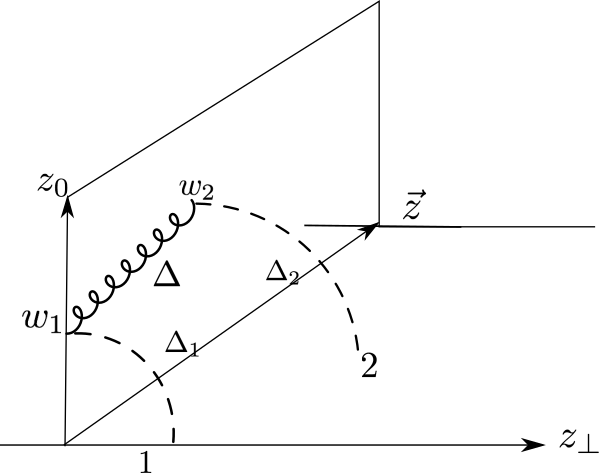}
\caption{The interface-channel geodesic Witten diagram.}
\label{fboundarygeo}
\end{center}
\end{figure}

Geodesic Witten diagrams are defined as modifications of the exchange Witten diagrams.  A geodesic Witten diagram with bulk exchange is represented by Figure \ref{fbulkgeo}. It differs from the bulk exchange Witten diagram (\ref{wdbulk}) by restricting the integral over the bulk point $Z$ to the geodesic $\gamma$ connecting the two inserted operators on the $AdS_{d+1}$ boundary. We denote it as
\begin{equation}\label{geowdbulk}
\mathcal{W}_{\rm bulk}(P_1,P_2)=\int_{\gamma}dZ\int_{AdS_d}dW G_{B\partial}^{\Delta_1}(P_1,Z)G_{B\partial}^{\Delta_2}(P_2,Z)G_{BB}^{\Delta}(Z,W)\;.
\end{equation}
A geodesic Witten diagram with interface exchange is represented by Figure \ref{fboundarygeo}. Instead of integrating over $W_1$ and $W_2$ as in (\ref{wdboundary}), $W_{1,2}$ are fixed by the coordinates $P_{1,2}$ of  the boundary insertions
\begin{equation}\label{fixcoord}
w_{i,0}=x_{i,\perp}\;,\;\;\;\;\;\;\;\;\;\;\;\;\; \vec{w}_i=\vec{x}_i\;,\;\;\;\;\;\;\;\;\;\;\;\;\; \vec{w}_\perp=0\;.
\end{equation}
The fixed points $W_i$ can be visualized as the intersection point of the $AdS_d$ brane with a geodesic line that emanates from $P_i$ and restricted to the $(z_0,z_\perp)$-plane such that the length of the geodesic is the shortest. We define this diagram as 
\begin{equation}\label{geowdboundary}
\mathcal{W}_{\rm interface}(P_1,P_2)=G_{B\partial}^{\Delta_1}(P_1,W_1)G_{BB}^{\Delta}(W_1,W_2)G_{B\partial}^{\Delta_2}(P_2,W_2)\;.
\end{equation}

In the next two subsections we will prove that the bulk and interface geodesic Witten diagrams (\ref{geowdbulk}) (\ref{geowdboundary}) are equivalent to the bulk and interface channel conformal blocks (\ref{bulkcb}) (\ref{boundarycb}), respectively. While the prescription of the bulk geodesic Witten diagram is clearly reminiscent to the work \cite{Hijano:2015zsa}, our prescription for the interface case is less obvious. We provide here a motivation using a heuristic argument used in \cite{Hijano:2015zsa}. The argument involves taking the limit where the external dimensions $\Delta_{1,2}$ are large while keeping $\Delta$ fixed. When the interface exchange Witten diagram is decomposed into interface-channel conformal blocks in this limit, the leading contribution is a conformal block that corresponds to the exchanged single-trace operator with dimension $\Delta$. Other conformal blocks would have dimensions $\Delta_{1,2}$ or higher and are exponentially suppressed. On the other hand, in the Witten diagram the integration over $W_{1,2}$ localizes as the fields dual to $O_{\Delta_{1,2}}$ become heavy. Indeed, most contribution from the bulk-to-boundary propagators $G^{\Delta_{1,2}}_{B\partial}(P_i,W_i)$ is localized at 
\begin{equation}
\begin{split}
{}&\frac{d}{dw_{i,0}}G^{\Delta_{i}}_{B\partial}=\frac{d}{dw_{i,0}}\left(\frac{w_{i,0}}{w_{i,0}^2+x_{i,\perp}^2+(\vec{w}_i-\vec{x})^2}\right)^{\Delta_i}=0\;,\\
{}&\frac{d}{d\vec{w}_i}G^{\Delta_{i}}_{B\partial}=\frac{d}{d\vec{w}_i}\left(\frac{w_{i,0}}{w_{i,0}^2+x_{i,\perp}^2+(\vec{w}_i-\vec{x})^2}\right)^{\Delta_i}=0\;.
\end{split}
\end{equation} 
The solution to these equations is just the prescription (\ref{fixcoord}). Therefore the heavy limit suggests us to identify $\mathcal{W}_{\rm interface}$ with the interface- channel conformal block.

\subsection{Proof by conformal Casimir equation}

\label{sub3}
We now give a first proof of the equivalence, using the conformal Casimir equation. 

Let us first focus on the bulk case. It is obvious that the integral 
\begin{equation}
\mathcal{I}_{\rm bulk}(W)=\int_{\gamma} dZ\;  G^{\Delta_1}_{B\partial}(P_1, Z)\;G^{\Delta_2}_{B\partial}(P_2, Z)\;G^{\Delta}_{BB}(Z, W)
\end{equation}
is invariant under the action of the operator
\begin{equation}
L_1^{AB}+L_2^{AB}+\mathcal{L}_W^{AB}\;.
\end{equation}
Here $L_{1,2}^{AB}$ represents the generators of conformal transformation  of boundary point 1 and 2. The operator $\mathcal{L}_W^{AB}$ is the $AdS_{d+1}$ isometry generator for the bulk point $W$. Hence applying the bulk-channel Casimir operator,
\begin{equation}
L^{2}_{\rm bulk}\; \mathcal{I}_{\rm bulk}(W)\equiv (L_1^{AB}+L_2^{AB})^2\mathcal{I}_{\rm bulk}(W)=(L_W^{AB})^2\mathcal{I}_{\rm bulk}(W)\;.
\end{equation}
Note that $(L_W^{AB})^2=-\square_{AdS_{d+1},W}$ and that $\square_{AdS_{d+1},W} G^{\Delta}_{BB}(Z, W)=\Delta(\Delta-d)G^{\Delta}_{BB}(Z, W)$. Using  $\mathcal{W}_{\rm bulk}=\int_{AdS_d} dW\mathcal{I}_{\rm bulk}$ we have thus shown that the bulk geodesic Witten diagram satisfies the Casimir equation
\begin{equation}
L^{2}_{\rm bulk}\;\mathcal{W}_{\rm bulk}=-\Delta(\Delta-d)\mathcal{W}_{\rm bulk}\;.
\end{equation}
Moreover, the  boundary conditions obeyed by the propagators guarantee that $\mathcal{W}_{\rm bulk}$ will have the correct power-law behavior $\mathcal{W}_{\rm bulk}\sim |x_{12}|^{\Delta-\Delta_1-\Delta_2}$ as we take the $x_1\to x_2$ limit. In summary, both $\mathcal{W}_{\rm bulk}$  the bulk-channel conformal block obey the same differential equation with the same boundary conditions, and are thus the same function.

Now let us move on to the interface case. Because of our prescription (\ref{fixcoord}), the bulk-to-boundary propagators are
\begin{equation}
G^{\Delta_i}_{B\partial}(P_i,W_i)=x_{i,\perp}^{-\Delta_i}
\end{equation}
and hence are not acted upon by the $SO(d,1)$ generators. On the other hand, the bulk-to-bulk propagator transforms under $SO(d,1)$ action of the residual conformal group thanks to (\ref{fixcoord}). Because the $SO(d,1)$ residual conformal group is the same as the isometry group of the $AdS_d$ brane, the action on  $G^{\Delta}_{BB}(W_1,W_2)$ can be written as
\begin{equation}
\hat{L}_1^{\hat{A}\hat{B}}\;G^{\Delta}_{BB}(W_1,W_2)=\hat{\mathcal{L}}_{W_1}^{\hat{A}\hat{B}}\;G^{\Delta}_{BB}(W_1,W_2)\, ,
\end{equation}
where $\hat{\mathcal{L}}_{W_1}^{\hat{A}\hat{B}}$ is the $AdS_d$ isometry generator. Using the relation twice and notice $(\hat{\mathcal{L}}_{W_1}^{\hat{A}\hat{B}})^2=-\square_{AdS_d,W_1}$ and $\square_{AdS_d,W_1}G^{\Delta}_{BB}(W_1,W_2)=\Delta(\Delta-d+1)G^{\Delta}_{BB}(W_1,W_2)$, we get
\begin{equation}
L^{2}_{\rm interface}\;\mathcal{W}_{\rm interface}\equiv(\hat{L}_1^{\hat{A}\hat{B}})^2\mathcal{W}_{\rm interface}=-\Delta(\Delta-d+1)\mathcal{W}_{\rm interface}\;.
\end{equation}
We have shown that $\mathcal{W}_{\rm interface}$ satisfies the interface Casimir equation. It is also easy to see that $\mathcal{W}_{\rm interface}$ satisfies the correct boundary condition. This concludes  the proof
that  $\mathcal{W}_{\rm interface}$ coincides the interface-channel conformal block.

\subsection{Proof by explicit evaluation}
\label{sub4}
We provide here another proof by explicit evaluation of the integrals. Casual readers may skip this subsection. The interface geodesic Witten diagram is equivalent to the interface conformal block by inspection
(compare (\ref{blkblkprop}) with (\ref{boundarycb})), so we will only consider the bulk case.

The geodesic represented as the dashed curve in Figure \ref{fbulkgeo} is a semicircle in  Poincar\'e coordinates. Using the AdS isometry, we can place the two boundary points on a plane spanned by the $z_{d-1}$ and $z_\perp$ direction so that the semicircle geodesic dips into the radial $z_0$ direction perpendicular to the $(z_{d-1},z_\perp)$-plane. It is convenient to introduce complex coordinates on this plane
\begin{equation}
\zeta=z_{d-1}+iz_\perp\;,\quad \bar{\zeta}=z_{d-1}-iz_\perp\;.
\end{equation}
Then the semicircle can be described in terms of complex coordinates as
\begin{equation}
2z_0^2+(\zeta-\zeta_1)(\bar{\zeta}-\bar{\zeta}_2)+(\bar{\zeta}-\bar{\zeta}_1)(\zeta-\zeta_2)=0\;.
\end{equation}
Here $\zeta_1$ and $\zeta_2$ are the complex coordinates of point 1 and 2 and we define $\alpha=x_{2,\perp}/x_{1,\perp}$. Any point on this semicircle is parameterized by the proper length $\lambda$
\begin{equation}
\begin{split}
\zeta(\lambda)={}&\frac{\zeta_1+\zeta_2}{2}+\frac{\zeta_2-\zeta_1}{2}\tanh\lambda\;,\\
\bar{\zeta}(\lambda)={}&\frac{\bar{\zeta}_1+\bar{\zeta}_2}{2}+\frac{\bar{\zeta}_2-\bar{\zeta}_1}{2}\tanh\lambda\;,\\
z_0(\lambda)={}&\frac{|\zeta_1-\zeta_2|}{2\cosh\lambda}\;.
\end{split}
\end{equation}
 For an arbitrary point labeled by $\lambda$ on this semicircle, the bulk-to-boundary propagator is very simple:
\begin{equation}
\begin{split}
G^{\Delta_1}_{B\partial}={}&\left(\frac{\frac{|\zeta_1-\zeta_2|}{2\cosh\lambda}}{\frac{|\zeta_1-\zeta_2|^2}{(2\cosh\lambda)^2}+\left|\frac{|\zeta_1-\zeta_2|}{2}\right|^2(1+\tanh\lambda)^2}\right)^{\Delta_1}=|\zeta_1-\zeta_2|^{-\Delta_1}e^{-\lambda \Delta_1}\;,\\
G^{\Delta_2}_{B\partial}={}&\left(\frac{\frac{|\zeta_1-\zeta_2|}{2\cosh\lambda}}{\frac{|\zeta_1-\zeta_2|^2}{(2\cosh\lambda)^2}+\left|\frac{|\zeta_1-\zeta_2|}{2}\right|^2(1-\tanh\lambda)^2}\right)^{\Delta_2}=|\zeta_1-\zeta_2|^{-\Delta_2}e^{+\lambda \Delta_2}\;.
\end{split}
\end{equation}
To handle the bulk-to-bulk propagator $G^{\Delta}_{BB}$, it turns out useful if we first apply a quadratic transformation on the hypergeometric function
\begin{equation}
G_{BB}^\Delta(Z,W)=u^\Delta\; {}_2F_1(\frac{\Delta}{2},\frac{\Delta}{2}+\frac{1}{2};\Delta-\frac{d}{2}+1;u^2)
\end{equation}
with a new variable $u=\frac{2w_0z_0}{w_0^2+z_0^2+z_\perp^2+(\vec{w}-\vec{z})^2}$. We then express the ${}_2F_1$ function using the Barnes representation 
\begin{equation}
{}_2F_1(a,b;c;z)=\frac{\Gamma(c)}{\Gamma(a)\Gamma(b)}\int_{-i\infty}^{+i\infty} d\tau \frac{\Gamma(-\tau)\Gamma(\tau+a)\Gamma(\tau+b)}{\Gamma(\tau+c)}(-z)^\tau \, .
\end{equation}
Note that here and in what follows, to avoid keeping writing the factor $(2\pi i)^{-1}$ we have absorbed it into the definition of the contour integral. We will then use the following elementary  integral identity
\begin{equation}
\int \frac{dw_0d\vec{w}}{w_0^d}\left(\frac{2w_0z_0}{w_0^2+z_0^2+z_\perp^2+(\vec{w}-\vec{z})^2}\right)^a=\frac{\pi^{\frac{d-1}{2}}\Gamma(\frac{a-d+1}{2})\Gamma(\frac{a}{2})(2z_0)^a}{2\Gamma(a)(z_0^2+z_\perp^2)^{\frac{a}{2}}}
\end{equation}
and the Barnes representation of ${}_2F_1$ to perform the $W$ integral. Without too much trouble we can derive
\begin{equation}\begin{split}
\int \frac{dw_0d\vec{w}}{w_0^d} G_{BB}^\Delta(Z,W)={}&\frac{ \pi^{\frac{d-1}{2}} \Gamma(\Delta-\frac{d}{2}-1)}{\Gamma(\frac{\Delta}{2})\Gamma(\frac{\Delta+1}{2})}\int_{-i\infty}^{+i\infty} d\tau (-1)^\tau 2^{2\tau+\Delta-1}\left(\frac{z_0^2}{z_0+z_\perp^2}\right)^{\tau+\frac{\Delta}{2}}\\
{}&\times\frac{\Gamma(-\tau)\Gamma(\tau+\frac{\Delta}{2})\Gamma(\tau+\frac{\Delta}{2}+1)}{\Gamma(\tau+\Delta-\frac{d}{2}+1)}\times \frac{\Gamma(\tau+\frac{\Delta+d-1}{2})\Gamma(\tau+\frac{\Delta}{2})}{\Gamma(2\tau+\Delta)}\;.
\end{split}\end{equation} 
After simplifying the expression using the duplication formula of Gamma function, we can exploit the Barnes representation to write the result as
\begin{equation}\label{gwdeq1}
\int \frac{dw_0d\vec{w}}{w_0^d} G_{BB}^\Delta=\frac{ \pi^{\frac{d}{2}} \Gamma(\frac{\Delta-d-1}{2})}{\Gamma(\frac{\Delta+1}{2})} \gamma^{\frac{\Delta}{2}}{}_2F_1(\frac{\Delta}{2},\frac{\Delta-d+1}{2};\Delta-\frac{d}{2}+1;\gamma)
\end{equation}
where 
\begin{equation}
\gamma=\frac{z_0^2}{z_0+z_\perp^2}\;.
\end{equation}
To proceed we must use the following hypergeometric identity,
\begin{equation}
{}_2F_1(a,b;c;z)=(1-z)^{-a}{}_2F_1(a,c-b;c;\frac{z}{z-1})\;.
\end{equation}
Then (\ref{gwdeq1})  becomes
\begin{equation}
\frac{ \pi^{\frac{d}{2}} \Gamma(\frac{\Delta-d-1}{2})}{\Gamma(\frac{\Delta+1}{2})} \left(\frac{z_0^2}{z_\perp^2}\right)^{\frac{\Delta}{2}}{}_2F_1(\frac{\Delta}{2},\frac{\Delta}{2}+\frac{1}{2};\Delta-\frac{d}{2}+1;-\frac{z_0^2}{z_\perp^2})\,.
\end{equation}
All in all, we have shown that the geodesic Witten diagram is given by the following integral  representation,
\begin{equation}
\mathcal{W}_{\rm bulk}=C\int_{-\infty}^{\infty}d\lambda |z_1-z_2|^{-(\Delta_1+\Delta_2)}e^{-\lambda(\Delta_1-\Delta_2)}\left(\frac{z_0^2}{z_\perp^2}\right)^{\frac{\Delta}{2}}{}_2F_1(\frac{\Delta}{2},\frac{\Delta}{2}+\frac{1}{2};\Delta-\frac{d}{2}+1;-\frac{z_0^2}{z_\perp^2})
\end{equation}
where $C$ is some numerical factor we will not write down explicitly in the intermediate steps. We can evaluate this integral by making a convenient change of variable,
\begin{equation}
\sigma\equiv \frac{e^{2\lambda}}{1+e^{2\lambda}} \, ,
\end{equation}
under which
\begin{equation}
\frac{z_0^2}{z_\perp^2}=\frac{|z_1-z_2|^2}{x_\perp^2}\frac{\sigma(1-\sigma)}{\left((1-\sigma)+\alpha\sigma\right)^2}\;.
\end{equation}
With the Barnes representation of ${}_2F_1$, the geodesic Witten diagram simplifies into  the following manageable integral,
\begin{equation}
\begin{split}
\mathcal{W}_{\rm bulk}={}&\frac{C}{2}\int_0^1\frac{d\sigma}{\sigma(1-\sigma)}|z_1-z_2|^{-(\Delta_1+\Delta_2)} \int_{-i\infty}^{+i\infty} d\tau \sigma^{\tau+\frac{\Delta+\Delta_2-\Delta_1}{2}}(1-\sigma)^{\tau+\frac{\Delta+\Delta_1-\Delta_2}{2}}\\
{}&\times \left(\frac{|z_1-z_2|^2}{x_\perp^2}\right)^{\tau+\frac{\Delta}{2}}\frac{\Gamma(-\tau)\Gamma(\tau+\frac{\Delta}{2})\Gamma(\tau+\frac{\Delta}{2}+\frac{1}{2})}{\Gamma(\tau+\Delta-\frac{d}{2}+1)}\times \frac{1}{\left((1-\sigma)+\alpha\sigma\right)^{2\tau+\Delta}}\;.
\end{split}
\end{equation}
The $\sigma$-integral is just a Feynman parameterization and gives a beta-function. After using again the duplication formula to clean up the Gamma function factors, one recognizes the $\tau$-integral can be written into another ${}_2F_1$ function. The geodesic Witten diagram therefore evaluates to
\begin{equation}
\mathcal{W}_{\rm bulk}= \frac{C'}{x_\perp^{\Delta_1}(\alpha x_\perp)^{\Delta_2}}\left(\frac{\eta}{4}\right)^{\frac{\Delta-\Delta_1-\Delta_2}{2}}{}_2F_1(\frac{\Delta+\Delta_1-\Delta_2}{2},\frac{\Delta+\Delta_2-\Delta_1}{2};\Delta-\frac{d}{2}+1;-\frac{\eta}{4})\, ,
\end{equation}
with the constant
\begin{equation}
C'=\frac{\pi^{\frac{d-1}{2}}\Gamma(\frac{\Delta-d+1}{2})\Gamma(\frac{\Delta+\Delta_1-\Delta_2}{2})\Gamma(\frac{\Delta+\Delta_2-\Delta_1}{2})}{\Gamma(\frac{\Delta}{2})\Gamma(\frac{\Delta+1}{2})^2}\;.
\end{equation}
We recognize that this is just the bulk-channel conformal block.

\section{Contact Witten diagrams}\label{Wcontact}
In this section we  perform a systematic evaluation of contact Witten diagrams $W_{n, m}$ with $n$ bulk and $m$ interface insertions.
We first warm up with two two cases that do not have cross ratios and can  thus be fully integrated, and then proceed to the more interesting cases.

\subsection{Contact diagrams without cross ratio}\label{blkbdr2pt}
The  $W_{1, 0}$ contact diagram consists of a bulk-to-boundary propagator with the bulk point integrated over the $AdS_d$ subspace,
\begin{equation}
W_{1,0}(P)= \int_{AdS_d} dW (-2P\cdot W)^{-\Delta}=\int \frac{dw_0d\vec{w}}{w_0^d} \left(\frac{w_{0}}{w_{0}^2+x_{\perp}^2+(\vec{w}-\vec{x})^2}\right)^{\Delta}\;.
\end{equation}
One evaluates this integral by moving the denominator to the exponent using Schwinger parameterization and then perform the $AdS_d$ integration. The  $AdS_d$ integral is an elementary Gaussian integral and the result is
\begin{equation}
W_{1,0}(P)=\frac{\pi^{(d-1)/2}\Gamma(\frac{\Delta}{2})\Gamma(\frac{\Delta-(d-2)}{2})}{2x_\perp^{\Delta}\Gamma(\Delta)}\; ,
\end{equation}
in agreement with the expected kinematic form of the one-point function.

Next, we consider $W_{1, 1}$, which consists of two bulk-to-boundary propagators that share a common integrated-over bulk point. We have two boundary points where operators are inserted. One is located at the boundary of $AdS_{d+1}$ and the other one is at the boundary of $AdS_{d}$. The common bulk point is confined on $AdS_{d}$ and is integrated over. Explicitly, 
\begin{equation}
\begin{split}
W_{1,1}(P,\hat{P})={}&\langle O(x)_{\Delta_1}\widehat{O}_{\widehat{\Delta}{_2}}(\vec{y})\rangle_{\rm contact}=\int_{AdS_d} dW (-2P\cdot W)^{\Delta_1}(-2\widehat{P}\cdot W)^{\widehat{\Delta}_2}\\
={}& \int \frac{dw_0d\vec{w}}{w_0^d} \left(\frac{w_{0}}{w_{0}^2+x_{\perp}^2+(\vec{w}-\vec{x})^2}\right)^{\Delta_1}\left(\frac{w_{0}}{w_{0}^2+(\vec{w}-\vec{y})^2}\right)^{\widehat{\Delta}_2}\;.
\end{split}
\end{equation}
The strategy of evaluation is similar to the previous case: we apply Schwinger parameterization twice to bring both denominators into the exponent and then integrate over AdS. The AdS integral is still elementary and we get
\begin{equation}
\frac{\pi^{(d-1)/2}\Gamma(\frac{\Delta_1+\widehat{\Delta}_2-(d-1)}{2})}{2\Gamma(\Delta_1)\Gamma(\widehat{\Delta}_2)}\int_0^\infty\frac{ds}{s}\frac{dt}{t}s^{\Delta_1}t^{\widehat{\Delta}_2}(s+t)^{-\frac{\Delta_1+\widehat{\Delta}_2}{2}}\times \exp(-sx_\perp^2-\frac{st(\vec{x}-\vec{y})^2}{s+t}) \;.
\end{equation}
To proceed, we  insert $1=\int_0^\infty d\rho\; \delta(\rho-s-t)$ into the integral so that
\begin{equation}
\frac{\pi^{(d-1)/2}\Gamma(\frac{\Delta_1+\widehat{\Delta}_2-(d-1)}{2})}{2\Gamma(\Delta_1)\Gamma(\widehat{\Delta}_2)}\int_0^\infty\frac{ds}{s}\frac{dt}{t}s^{\Delta_1}t^{\widehat{\Delta}_2}\int_0^\infty d\rho\; \delta(\rho-s-t) \rho^{-\frac{\Delta_1+\widehat{\Delta}_2}{2}}e^{-sx_\perp^2-\frac{st(\vec{x}-\vec{y})^2}{\rho}}.
\end{equation}
Then we rescale $s\to \rho s$, $t\to \rho t$. Notice the delta function becomes $\delta(\rho(1-s-t))=\rho^{-1}\delta(1-s-t)$ and therefore restricts the integration region of both $s$ and $t$ to $[0,1]$,
\begin{equation}\label{W11}
\begin{split}
{}&W_{1,1}(P,\hat{P})\\
{}&=\frac{\pi^{\frac{d-1}{2}}\Gamma(\frac{\Delta_1+\widehat{\Delta}_2-(d-1)}{2})}{2\Gamma(\Delta_1)\Gamma(\widehat{\Delta}_2)}\int_0^1\frac{ds}{s}\frac{dt}{t}s^{\Delta_1}t^{\widehat{\Delta}_2}\int_0^\infty \frac{d\rho}{\rho}\; \delta(1-s-t) \rho^{\frac{\Delta_1+\widehat{\Delta}_2}{2}} e^{-\rho sx_\perp^2-\rho{st(\vec{x}-\vec{y})^2}}\\
{}&= \frac{\pi^{\frac{d-1}{2}}\Gamma(\frac{\Delta_1+\widehat{\Delta}_2-(d-1)}{2})\Gamma(\frac{\Delta_1+\widehat{\Delta}_2}{2})}{2\Gamma(\Delta_1)\Gamma(\widehat{\Delta}_2)}\int_0^1\frac{ds}{s}\frac{dt}{t}\frac{\delta(1-s-t)s^{\frac{\Delta_1-\Delta_2}{2}}t^{\widehat{\Delta}_2}}{(s x_\perp^2+t((\vec{x}-\vec{y})^2+x_\perp^2))^{\frac{\Delta_1-\widehat{\Delta}_2}{2}}}\\
{}&=\frac{\pi^{\frac{d-1}{2}}\Gamma(\frac{\Delta_1+\widehat{\Delta}_2-(d-1)}{2})\Gamma(\frac{\Delta_1-\widehat{\Delta}_2}{2})}{2\Gamma(\Delta_1)}\frac{1}{x_\perp^{\Delta_1-\widehat{\Delta}_2}(x_\perp^2+(\vec{x}-\vec{y})^2)^{\widehat{\Delta}_2}}\;. 
\end{split}
\end{equation}
In the last step, we have used Feynman's parameterization for the integral involving $s$ and $t$. This result is again consistent with the kinematic expectation: there is only one invariant with the correct scaling, namely
\begin{equation}
(-2P_1\cdot \widehat{P}_2)^{-\widehat{\Delta}_2}(P_1\cdot B)^{\widehat{\Delta}_2-\Delta_1}=\frac{1}{x_\perp^{\Delta_1-\widehat{\Delta}_2}(x_\perp^2+(\vec{x}-\vec{y})^2)^{\widehat{\Delta}_2}}\;.
\end{equation}

\subsection{Contact diagrams with one cross ratio}
Now we are ready to use the same basic tricks to evaluate the two-point contact Witten diagram in Figure \ref{fcontact},
\begin{equation}
\begin{split}
{}&W_{2,0}(P_1,P_2)=\langle O_{\Delta_1}(x)O_{\Delta_2}(y) \rangle_{\rm contact}=\int_{AdS_d}dW G^{\Delta_1}_{\partial B}(P_1,W)G^{\Delta_2}_{\partial B}(P_2,W)\\
{}&=\int \frac{dw_0d\vec{w}}{w_0^d}\left(\frac{w_0}{w_0^2+x_\perp^2+(\vec{w}-\vec{x})^2}\right)^{\Delta_1}\left(\frac{w_0}{w_0^2+y_\perp^2+(\vec{w}-\vec{y})^2}\right)^{\Delta_2}\\
{}&=\frac{\pi^{\frac{d-1}{2}}\Gamma(\frac{\Delta_1+\Delta_2-(d-1)}{2})}{2\Gamma(\Delta_1)\Gamma(\Delta_2)}\int \frac{ds}{s}\frac{dt}{t}s^{\Delta_1}t^{\Delta_2}(s+t)^{-\frac{\Delta_1+\Delta_2}{2}}e^{-sx_\perp^2-ty_\perp^2-\frac{st (\vec{x}-\vec{y})^2}{s+t}}\;.
\end{split}
\end{equation}
In the above we have already used Schwinger's trick for the denominators. It is important to organize the exponent in the last line into
\begin{equation}
-\frac{st (x-y)^2}{s+t}-\frac{(sx_\perp+ty_\perp)^2}{s+t}\, ,
\end{equation}
where we remind the reader that $(x-y)^2$ is our short-hand notation for $(\vec{x}-\vec{y})^2+(x_\perp-y_\perp)^2$. Inserting again $1=\int_0^\infty d\rho\; \delta(\rho-s-t)$ and rescale $s$, $t$ by $\rho$, the integral becomes
\begin{equation}
\begin{split}
W_{2,0}={}&\frac{\pi^{\frac{d-1}{2}}\Gamma(\frac{\Delta_1+\Delta_2-(d-1)}{2})}{2\Gamma(\Delta_1)\Gamma(\Delta_2)}\int_0^1 \frac{ds}{s}\frac{dt}{t}s^{\Delta_1}t^{\Delta_2}\;\delta(1-s-t)\\
\times{}&\int_0^\infty \frac{d\rho}{\rho}\rho^{\frac{\Delta_1+\Delta_2}{2}}e^{-\rho st (x-y)^2-\rho (sx_\perp+ty_\perp)^2}\;.
\end{split}
\end{equation}
We use the inverse Mellin transformation of $e^{-z}$,
\begin{equation}
e^{-z}=\int_{-i\infty}^{i\infty} \Gamma(\tau)z^{-\tau} d\tau
\end{equation}
to represent the exponent $-\rho st (x-y)^2$ and the $\rho$ integral will just be of the familiar type $\int_0^\infty d\rho \rho^{\alpha-1} e^{-\rho}$. The manipulations lead to an expression which involves an integral over $\tau$ and two integrals over $s$ and $t$
\begin{equation}
\begin{split}
\int\displaylimits_{-i\infty}^{+i\infty} \frac{ d\tau\pi^{\frac{d-1}{2}}\Gamma(\frac{\Delta_1+\Delta_2-d+1)}{2})\Gamma(\tau)\Gamma(\frac{\Delta_1+\Delta_2-2\tau}{2})}{2\Gamma(\Delta_1)\Gamma(\Delta_2)}\int_0^1 \frac{dsdt\delta(1-s-t)s^{\Delta_1-\tau-1}t^{\Delta_2-\tau-1}}{(s x_\perp+ t y_\perp)^{\Delta_1+\Delta_2-2\tau}(x-y)^{2\tau}}.
\end{split}
\end{equation}
The second integral with $s$ and $t$ is again the Feynman parameterization and can be readily evaluated, yielding a $\tau$-integral
\begin{equation}\label{2ptcontact}
\frac{1}{x_\perp^{\Delta_1}y_\perp^{\Delta_2}}\frac{\pi^{\frac{d-1}{2}}\Gamma(\frac{\Delta_1+\Delta_2-(d-1)}{2})}{2\Gamma(\Delta_1)\Gamma(\Delta_2)}\int_{-i\infty}^{+i\infty} d\tau \frac{\Gamma(\Delta_1-\tau)\Gamma(\Delta_2-\tau)\Gamma(\tau)\Gamma(\frac{\Delta_1+\Delta_2}{2}-\tau)}{\Gamma(\Delta_1+\Delta_2-2\tau)}\eta^{-\tau}\;.
\end{equation}
This integral can be further tidied up using the duplication formula of Gamma function and it equals
\begin{equation}
\frac{1}{(2x_\perp)^{\Delta_1}(2y_\perp)^{\Delta_2}}\frac{\pi^{\frac{d}{2}}\Gamma(\frac{\Delta_1+\Delta_2-(d-1)}{2})}{\Gamma(\Delta_1)\Gamma(\Delta_2)}\int_{-i\infty}^{+i\infty} d\tau \frac{\Gamma(\Delta_1-\tau)\Gamma(\Delta_2-\tau)\Gamma(\tau)}{\Gamma(\frac{\Delta_1+\Delta_2+1}{2}-\tau)}\left(\frac{\eta}{4}\right)^{-\tau}\;,
\end{equation}
where $\eta$ is the cross ratio we defined in (\ref{defeta}). Comparing with the definition of Mellin amplitude (\ref{defmellin}), we find the Mellin amplitude of a two-point contact Witten diagram is just a constant,
\begin{equation}\label{mellincontact}
\mathcal{M}_{\rm contact}=\frac{\pi^{\frac{d}{2}}\Gamma(\frac{\Delta_1+\Delta_2-(d-1)}{2})}{\Gamma(\Delta_1)\Gamma(\Delta_2)}\;.
\end{equation}
We also point out here that the above integral is nothing but a ${}_2F_1$ function, therefore\footnote{{This two-point contact Witten diagram was also evaluated in \cite{Aharony:2003qf} using a different method. One can show that their result equation (79) is equivalent to (\ref{W202F1}) using the  hypergeometric identities
\begin{equation}
\nonumber {}_2F_1(a,b;2b;z)=(1-z)^{-\frac{a}{2}}{}_2F_1(\frac{a}{2},b-\frac{a}{2};b+\frac{1}{2};\frac{z^2}{4z-4})\;,
\end{equation}
\begin{equation}
\nonumber {}_2F_1(a,b;a+b+\frac{1}{2};z)={}_2F_1(2a,2b;a+b+\frac{1}{2};\frac{1-\sqrt{1-z}}{2})\;,
\end{equation}
and $\xi \equiv \eta/4$ to relate the different definitions of the cross ratio.
 }},
\begin{equation}\label{W202F1}
W_{2,0}(P_1,P_2)=\frac{\pi^{\frac{d}{2}}}{(2x_\perp)^{\Delta_1}(2y_\perp)^{\Delta_2}}\frac{\Gamma(\frac{\Delta_1+\Delta_2-(d-1)}{2})}{\Gamma(\frac{\Delta_1+\Delta_2+1}{2})}{}_2F_1(\Delta_1,\Delta_2;\frac{\Delta_1+\Delta_2+1}{2};-\frac{\eta}{4})\;.
\end{equation}
This two-point Witten diagram is in some sense the building blocks of other Witten diagrams. We will see in the next section, under the conspiracy of spectrum and spacetime dimension, exchange Witten diagrams can be evaluated as a finite sum of such contact diagrams. 

Equipped with the above techniques, one can calculate without too much effort another type of Witten diagrams that also have only one cross ratio. This is the case where we have two operators on the interface and one operator in the bulk. We do not repeat the intermediate steps here, but just state the result,
\begin{equation}
\begin{split}
W_{1,2}(P,\hat{P}_1,\hat{P}_2)\equiv{}&\langle O_{\Delta}(x)\widehat{O}_{\widehat{\Delta}{_1}}(\vec{x}_1)\widehat{O}_{\widehat{\Delta}{_2}}(\vec{x}_2)\rangle_{\rm contact}\\
={}&\int_{AdS_d}dW (-2P\cdot W)^{-\Delta}(-2\widehat{P}_1\cdot W)^{-\widehat{\Delta}_1}(-2\widehat{P}_2\cdot W)^{-\widehat{\Delta}_2}\\
={}&\frac{\pi^{(d-1)/2}\Gamma(\frac{\Delta+\widehat{\Delta}_1+\widehat{\Delta}_2-d+1}{2})}{2\Gamma(\Delta)\Gamma(\widehat{\Delta}_1)\Gamma(\widehat{\Delta}_2)}\frac{(P\cdot B)^{\widehat{\Delta}_1+\widehat{\Delta}_2-\Delta}}{(-2P\cdot \widehat{P}_1)^{\widehat{\Delta}_1}(-2P\cdot \widehat{P}_2)^{\widehat{\Delta}_2}}\\
\times{}& \int d\tau \Gamma(\tau)\Gamma(\widehat{\Delta}_1-\tau)\Gamma(\widehat{\Delta}_2-\tau)\Gamma(\frac{\Delta-\widehat{\Delta}_1-\widehat{\Delta}_2}{2}+\tau)\xi^{\tau}\, ,
\end{split}
\end{equation}
where
\begin{equation}
\xi=\frac{(-2P\cdot \widehat{P}_1)(-2P\cdot \widehat{P}_2)}{(-2\widehat{P}_1\cdot \widehat{P}_2) (P\cdot B)^2}
\end{equation} 
is the unique scaling-invariant cross ratio. 

We are now going to check our claim that its Mellin amplitude is a constant.  To this end we write the three-point function using the auxiliary momenta
\begin{equation}
\begin{split}
W_{1,2}(P,\hat{P}_1,\hat{P}_2)={}&\int d[\widehat{p}_i,p] (-2\widehat{P}_1\cdot\widehat{P}_2)^{-\widehat{p}_1\cdot\widehat{p}_2}(-2\widehat{P}_1\cdot{P})^{-\widehat{p}_1\cdot{p}}(-2\widehat{P}_2\cdot{P})^{-\widehat{p}_2\cdot{p}}\\
\times{}& (P\cdot B)^{-p\cdot \mathcal{P}}  \frac{\Gamma(\widehat{p}_1\cdot\widehat{p}_2)\Gamma(\widehat{p}_1\cdot{p})\Gamma(\widehat{p}_2\cdot{p})\Gamma({p}\cdot\mathcal{P})\Gamma(-\mathcal{P}^2)}{\Gamma(-\frac{\mathcal{P}^2}{2})}\mathcal{M}(\widehat{p}_i,p)\;.
\end{split}
\end{equation}
By inspection, the relation between the bilinears and $\tau$ is 
\begin{equation}
\begin{split}
\widehat{p}_1\cdot\widehat{p}_2={}&\frac{-\mathcal{P}^2-\Delta+\widehat{\Delta}_1+\widehat{\Delta}_2}{2}=\tau\;,\\
p\cdot \mathcal{P}={}&-\mathcal{P}^2=\Delta-\widehat{\Delta}_1-\widehat{\Delta}_2+2\tau\;,\\
{p}\cdot\widehat{p}_1={}&\frac{\mathcal{P}^2+\Delta+\widehat{\Delta}_1-\widehat{\Delta}_2}{2}=\widehat{\Delta}_1-\tau\;,\\
{p}\cdot\widehat{p}_2={}&\frac{\mathcal{P}^2+\Delta-\widehat{\Delta}_1+\widehat{\Delta}_2}{2}=\widehat{\Delta}_2-\tau\;,\\
\mathcal{P}^2={}&-2\tau-\Delta+\widehat{\Delta}_1+\widehat{\Delta}_2\;.
\end{split}
\end{equation}
Substituting these bilinears into Gamma functions in the definition, we happily find again that the Mellin amplitude for this three-point function is just a constant.

\subsection{General cases}
In this subsection we further generalize the observation that contact Witten diagrams have constant Mellin amplitudes. We consider contact Witten diagram with $n$ operators in the bulk and $m$ operators on the interface. In the following calculation, we only assume $n\geq 1$ since otherwise the calculation reduces to the known case where an interface is absent. Such a contact Witten diagram was denoted as $W_{n,m}$ and it is the following integral
\begin{equation}
\begin{split}
{}&W_{n,m} \equiv\langle O_{\Delta_1}(x_1)\ldots O_{\Delta_n}(x_n) \widehat{O}_{\widehat{\Delta}_1}(y_1)\ldots \widehat{O}_{\widehat{\Delta}_m}(y_m) \rangle_{\rm contact}\\
{}&=\int\frac{dw_0d\vec{w}}{w_0^d}\prod_{i}\left(\frac{w_0}{w_0^2+x_{\perp,i}^2+(\vec{w}-\vec{x}_i)^2}\right)^{\Delta_i}\prod_{I}\left(\frac{w_0}{w_0^2+(\vec{w}-\vec{y}_I)^2}\right)^{\widehat{\Delta}_I}\;.
\end{split}
\end{equation}
We use Schwinger's trick to bring the denominators into the exponent and perform the integrals of $dw_0$ and $d\vec{w}$, which leads to
\begin{equation}
\begin{split}
{}&W_{n,m}= \frac{\pi^{\frac{d-1}{2}}\Gamma(\frac{\sum_i \Delta_i+\sum_I\widehat{\Delta}_I-d+1}{2})}{2\prod_i\Gamma(\Delta_i)\prod_I\Gamma(\widehat{\Delta}_I)}\int_0^\infty\prod_i\frac{dt_i}{t_i}t_i^{\Delta_i}\frac{ds_I}{s_I}s_I^{\widehat{\Delta}_I} (\sum_i t_i+\sum_I s_I)^{-\frac{\sum_i \Delta_i+\sum_I\widehat{\Delta}_I}{2}}\\
{}&\times \exp\left[-\frac{\sum\limits_{i<j}t_it_j(-2P_i\cdot P_j)+(\sum\limits_it_iP_i\cdot B)^2+\sum\limits_{I<J}s_Is_J(-2 \widehat{P}_I\cdot \widehat{P}_J)+\sum\limits_{i,I}t_is_I(-2P_i\cdot \widehat{P}_I)}{\sum\limits_i t_i+\sum\limits_Is_I}\right].
\end{split}
\end{equation}
To proceed, we insert 
\begin{equation}
\int_0^{\infty}d\rho\;\delta(\rho-\sum_it_i-\sum_Is_I)=1
\end{equation}
to replace all $\sum_i t_i+\sum_Is_I$ by $\rho$. Then we rescale $t_i$ and $s_I$ by $\rho^{1/2}$ so that all the powers of $\rho$ are removed. Notice, after the rescaling, the only integral in $\rho$ is the following delta function
\begin{equation}
\int_0^{\infty}d\rho\;\delta(\rho-\sqrt{\rho}(\sum_it_i+\sum_Is_I))=2\;.
\end{equation}
This turns the integral into
\begin{equation}
\begin{split}
{}&W_{n,m}= \frac{\pi^{\frac{d-1}{2}}\Gamma(\frac{\sum_i \Delta_i+\sum_I\widehat{\Delta}_I-d+1}{2})}{\prod_i\Gamma(\Delta_i)\prod_I\Gamma(\widehat{\Delta}_I)}\int_0^\infty\prod_i\frac{dt_i}{t_i}t_i^{\Delta_i}\frac{ds_I}{s_I}s_I^{\widehat{\Delta}_I} \\
{}&\times \exp\left[-{\sum\limits_{i<j}t_it_j(-2P_i\cdot P_j)-(\sum\limits_it_iP_i\cdot B)^2-\sum\limits_{I<J}s_Is_J(-2 \widehat{P}_I\cdot \widehat{P}_J)-\sum\limits_{i,I}t_is_I(-2P_i\cdot \widehat{P}_I)}\right].
\end{split}
\end{equation}
We use the Mellin representation of exponential for the following terms in the exponent: all $t_it_j$, all $s_Is_J$ and the $t_is_I$ with $i>1$. Their conjugate variables are respectively denoted as $\delta_{ij}$, $\beta_{IJ}$ and $\gamma_{iI}$. We then get the following integral
\begin{equation} 
\label{generalW}
\begin{split}
W_{n,m} ={}&\frac{\pi^{\frac{d-1}{2}}\Gamma(\frac{\sum_i \Delta_i+\sum_I\widehat{\Delta}_I-d+1}{2})}{\prod_i\Gamma(\Delta_i)\prod_I\Gamma(\widehat{\Delta}_I)} \int \prod_{i<j}\left([d\delta_{ij}] \Gamma(\delta_{ij}) (-2 P_i\cdot P_j)^{-\delta_{ij}}\right)\\
{}&\times \int \prod_{I<J}\left([d\beta_{IJ}] \Gamma(\beta_{IJ}) (-2 \widehat{P}_I\cdot \widehat{P}_J)^{-\beta_{IJ}}\right) \prod_{i>1,I}\left([d\gamma_{iI}] \Gamma(\gamma_{iI}) (-2 P_i\cdot \widehat{P}_I)^{-\gamma_{iI}}\right)\\
{}&\times\int_0^\infty \prod_i \frac{dt_i}{t_i}t_i^{\Delta_i-\sum_{j\neq i}\delta_{ij}-\sum_I[1-\delta(i,1)]\gamma_{iI}}\int_0^\infty \prod_I \frac{ds_I}{s_I}s_I^{\widehat{\Delta}_I-\sum_{J\neq I}\beta_{IJ}-\sum_{i>1}\gamma_{iI}}\\
{}&\times \exp\left(-(\sum_it_iP_i\cdot B)^2\right)\exp\left(-\sum_{I}t_1s_I(-2P_1\cdot \widehat{P}_I)\right)\;.
\end{split}
\end{equation}
Here in the third line we denoted the Kronecker delta function as $\delta(a,b)$ in order to distinguish it from the Mellin variable $\delta_{ij}$, and we hope that it will not cause any confusion to the reader. The $s$-integral can now be easily performed, giving 
\begin{equation}
\prod_I \left[(-2P_1\cdot \widehat{P}_I)^{-\widehat{\Delta}_I+\sum_{J\neq I}\beta_{IJ}+\sum_{i>1}\gamma_{iI}}\Gamma(\widehat{\Delta}_I-\sum_{J\neq I}\beta_{IJ}-\sum_{i>1}\gamma_{iI})\right]t_1^{-\sum_I(\widehat{\Delta}_I-\sum_{J\neq I}\beta_{IJ}-\sum_{i>1}\gamma_{iI})}\;.
\end{equation}
We notice that the combination $\widehat{\Delta}_I-\sum_{J\neq I}\beta_{IJ}-\sum_{i>1}\gamma_{iI}$ is just $\gamma_{1I}$ by (\ref{constraints}). Plugging the $s$-integral result into the total integral, the $t$-integral just becomes 
\begin{equation}
\int_0^\infty \prod_{i=1}^n \frac{dt_i}{t_i} t_i^{\Delta_i-\sum_{j\neq i}\delta_{ij}-\sum_I\gamma_{iI}}\exp\left(-(\sum_i t_iP_i\cdot B)^2\right)
\end{equation}
where we have used $\gamma_{1I}\equiv \widehat{\Delta}_I-\sum_{J\neq I}\beta_{IJ}-\sum_{i>1}\gamma_{iI}$. We can evaluate this integral by inserting 
\begin{equation}
\int_0^\infty d\lambda\;\delta(\lambda-\sum_it_i)=1\;,
\end{equation}
and rescaling $t_i\to\lambda t_i$. Define $\alpha_i\equiv \Delta_i-\sum_{j\neq i}\delta_{ij}-\sum_I\gamma_{iI}$ as in (\ref{constraints}), the new integral is
\begin{equation}
\int_0^\infty \prod_{i=1}^n \frac{dt_i}{t_i} t_i^{\alpha_i}\delta(1-\sum_i t_i)\int_0^\infty\frac{d\lambda}{\lambda}\lambda^{\sum_i\alpha_i}\exp\left(-\lambda^2(\sum_i t_iP_i\cdot B)^2\right)\;,
\end{equation} 
and it is not difficult to find that this integral evaluates to
\begin{equation}
\frac{1}{2}\left(\prod_i \Gamma(\alpha_i)(P_i\cdot B)^{-\alpha_i}\right)\frac{\Gamma(\frac{\sum_i\alpha_i}{2})}{\Gamma(\sum_i\alpha_i)}\;.
\end{equation}
All in all, we have obtained the following result for the general contact Witten diagram
\begin{equation} 
\label{generalW}
\begin{split}
W_{n,m} ={}&\frac{\pi^{\frac{d-1}{2}}\Gamma(\frac{\sum_i \Delta_i+\sum_I\widehat{\Delta}_I-d+1}{2})}{2\prod_i\Gamma(\Delta_i)\prod_I\Gamma(\widehat{\Delta}_I)} \int \prod_{i<j}\left([d\delta_{ij}] \Gamma(\delta_{ij}) (-2 P_i\cdot P_j)^{-\delta_{ij}}\right)\\
{}&\times \int \prod_{I<J}\left([d\beta_{IJ}] \Gamma(\beta_{IJ}) (-2 \widehat{P}_I\cdot \widehat{P}_J)^{-\beta_{IJ}}\right) \prod_{iI}\left([d\gamma_{iI}] \Gamma(\gamma_{iI}) (-2 P_i\cdot \widehat{P}_I)^{-\gamma_{iI}}\right)\\
{}&\times\prod_i \left(d[\alpha_i]\Gamma(\alpha_i)(P_i\cdot B)^{-\alpha_i}\right)\frac{\Gamma(\frac{\sum_i\alpha_i}{2})}{\Gamma(\sum_i\alpha_i)}\;
\end{split}
\end{equation}
where the integration variable are subject to the constraints (\ref{constraints}). 

Now let us extract the Mellin amplitude. Thanks to the relation
\begin{equation}
\sum_i\alpha_i=\sum_i \Delta_i-\sum_{i,I}\gamma_{iI}-\sum_{i\neq j}\delta_{ij} =-\sum_{i,j}p_i\cdot p_j-\sum_{i,I} p_i\cdot \widehat{p}_I=-\mathcal{P}^2\;,
\end{equation}
the outstanding ratio of Gamma functions therefore is just the one that appears in the definition (\ref{defmellingeneral}). We thus find that such contact Witten diagrams all have constant Mellin amplitudes.

\section{Exchange Witten diagrams}\label{Wexchange}

In this section we evaluate exchange Witten diagrams with two external bulk scalar operators. We consider both bulk and interface exchanges. We develop both the {\it truncation} method, which applies to special spectra that  are
sometimes satisfied by AdS supergravity theories, and the  {\it spectral representation} method,  which applies to general spectra. Our main results are the formulae (\ref{blkmellintruc}) and (\ref{bdrmellintruc}) for the truncation method and the formulae (\ref{blkmellinspec}), (\ref{bdrmellinspec}) for the spectral representation method.

\subsection{Bulk exchange Witten diagrams: the truncation method}\label{blktrunc} 
\begin{figure}[t]
\begin{center}
\includegraphics[scale=0.5]{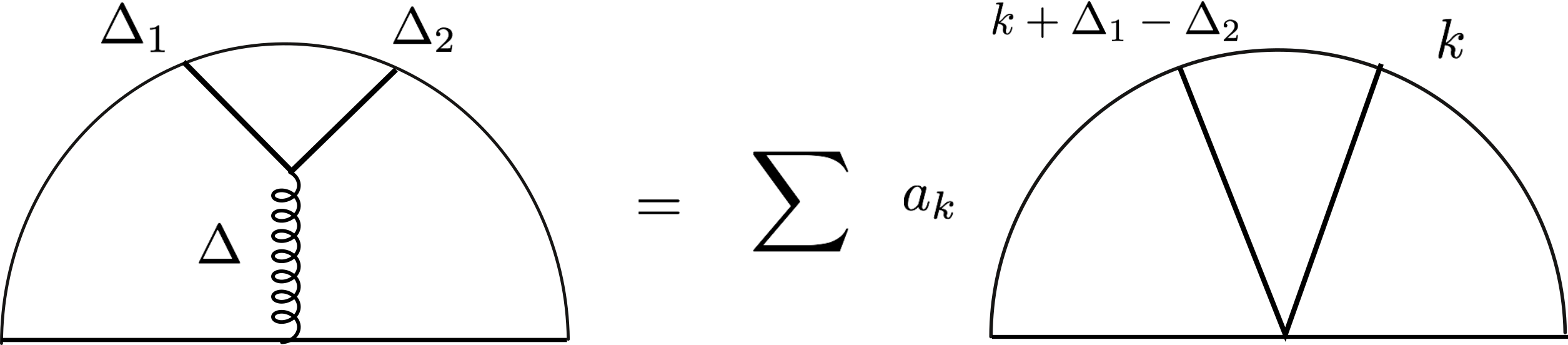}
\caption{A bulk exchange Witten diagram is replaced by a sum of contact Witten diagrams when $\Delta_1+\Delta_2-\Delta$ is a positive even integer.}
\label{fblktrunc}
\end{center}
\end{figure}

Let us start with the bulk exchange Witten diagram in Figure \ref{fbulkx}. The Witten diagram is given by the following integral,
\begin{equation}
W_{\rm bulk}= \int_{AdS_{d}} dW \int_{AdS_{d+1}} dZ\;  G^{\Delta_1}_{B\partial}(P_1, Z)\;G^{\Delta_2}_{B\partial}(P_2, Z)\;G^{\Delta}_{BB}(Z, W)\;.
\end{equation}
The $Z$-integral has been performed in \cite{DHoker:1999mqo} ``without really trying''.  Let us briefly review that method. Denote the $Z$-integral as
\begin{equation}
A(W,P_1,P_2)= \int_{AdS_{d+1}} dZ\;  G^{\Delta_1}_{B\partial}(P_1, Z)\;G^{\Delta_2}_{B\partial}(P_2, Z)\;G^{\Delta}_{BB}(Z, W).
\end{equation}
It is convenient to perform a translation such that
\begin{equation}
x_1\to 0\;,\;\;\;x_2\to x_{21}\equiv x_2-x_1\;.
\end{equation}
This is followed by a conformal inversion,
\begin{equation}
x_{12}'=\frac{x_{12}}{(x_{12})^2}\;,\;\;\; z'=\frac{z}{z^2}\;,\;\;\; w'=\frac{2}{w^2}.
\end{equation}
After these transformations the integral becomes,
\begin{equation}
A(W,P_1,P_2)=(x_{12})^{-2\Delta_2}I(w'-x_{12}')
\end{equation}
where
\begin{equation}
I(w)=\int\frac{d^{d+1}z}{z_0^{d+1}}\;G_{BB}^{\Delta}(-2Z\cdot W)\;z_0^{\Delta_1}\left(\frac{z_0}{z^2}\right)^{\Delta_2}\;.
\end{equation}
The scaling behavior of $I(w)$ under $w\to\lambda w$ together with the Poincar\'e symmetry dictates that $I(w)$ takes the form
\begin{equation}
I(w)=w_0^{\Delta_1-\Delta_2} f(t)
\end{equation}
where 
\begin{equation}
t=\frac{w_0^2}{w^2}
\end{equation}
On the other hand $f(t)$ is constrained by the following differential equation,
\begin{equation}
4t^2(t-1)f''+4t[(\Delta_1-\Delta_2+1)t-\Delta_1+\Delta_3+\frac{d}{2}-1]f'+[(\Delta_1-\Delta_2)(d-\Delta_1+\Delta_2)+m^2]f=t^{\Delta_2}
\end{equation}
where $m^2=\Delta(\Delta-d)$. This equation comes from acting with the equation of motion of the field in the bulk-to-bulk propagator,
\begin{equation}
-\square_{AdS_{d+1},W}G^{\Delta}_{BB}(Z,W)+m^2G^{\Delta}_{BB}(Z,W)=\delta(Z,W)
\end{equation}
and it collapses the bulk-to-bulk propagator to a delta-function. The solution to this equation is generically hypergeometric functions of type ${}_2F_1$ which expands to an infinite series, however with appropriate choice of conformal dimensions, $f(t)$ admits a polynomial solution:
\begin{equation}
f(t)=\sum_{k=k_{\rm min}}^{k_{\rm max}} a_k t^k
\end{equation} 
with
\begin{equation}
\begin{split}
{}&k_{\rm min}=(\Delta-\Delta_{1}+\Delta_2)/2\;,\;\;\;\;\; k_{\rm max}=\Delta_2-1\;,\\
{}&a_{k-1}=a_k \frac{(k-\frac{\Delta}{2}+\frac{\Delta_{1}-\Delta_2}{2})(k-\frac{d}{2}+\frac{\Delta}{2}+\frac{\Delta_{1}-\Delta_2}{2})}{(k-1)(k-1-\Delta_{1}+\Delta_2)}\;,\\
{}& a_{\Delta_2-1}=\frac{1}{4(\Delta_1-1)(\Delta_2-1)}
\end{split}
\end{equation}
and this truncation happens when $\Delta_1+\Delta_2-\Delta$ is a positive even integer. After obtaining this solution, we can undo the inversion and translation and the upshot is that $A(W,P_1,P_2)$ becomes a sum of contact vertices at $W$,
\begin{equation}
A(W,P_1,P_2)=\sum_{k=k_{\rm min}}^{k_{\rm max}} a_k (-2P_1\cdot P_2)^{k-\Delta_2} G_{B\partial}^{k+\Delta_1-\Delta_2}(P_1,W)\;G_{B\partial}^{k}(P_2,W)\;
\end{equation}
This identity is illustrated in Figure \ref{fblktrunc}.

We can then use the formula (\ref{2ptcontact}) for two-point contact Witten diagrams to obtain the Mellin amplitude of a bulk exchange Witten diagram,
\begin{equation}\label{blkmellintruc}
\mathcal{M}_{\rm bulk}=\pi^{d/2}\sum_{k_{\rm min}}^{k_{\rm max}}\frac{a_k \Gamma(\frac{\Delta_1-\Delta_2+2k-(d-1)}{2})}{\Gamma(\Delta_1-\Delta_2+k)\Gamma(k)}\frac{ \Gamma(\tau+k-\Delta_2)}{\Gamma(\tau)}\;.
\end{equation}
This Mellin amplitude has finitely many simple poles in $\tau$. Alternatively in terms of the squared interface momentum $\mathcal{P}^2=2\tau-\Delta_1-\Delta_2$, the simple poles of the Mellin amplitude are located at
\begin{equation}
-\Delta,-\Delta-2,\ldots,-\Delta_1-\Delta_2+2
\end{equation}
resembling a resonance amplitude with intermediate particles whose squared masses  are $\Delta,\Delta+2,\ldots,\Delta_1+\Delta_2-2$. 
\subsection{Bulk exchange Witten diagrams: the spectral representation}\label{blkspec}
The method used in the previous section may work nicely in a Kaluza-Klein supergravity theory with an integer-spaced spectrum of conformal dimensions. Generically such truncation need not take place and we must resort to a more general method to evaluate the Witten diagrams. In this subsection, we will evaluate the bulk exchange Witten diagrams using spectral representations.

\begin{figure}[t]
\begin{center}
\includegraphics[scale=0.6]{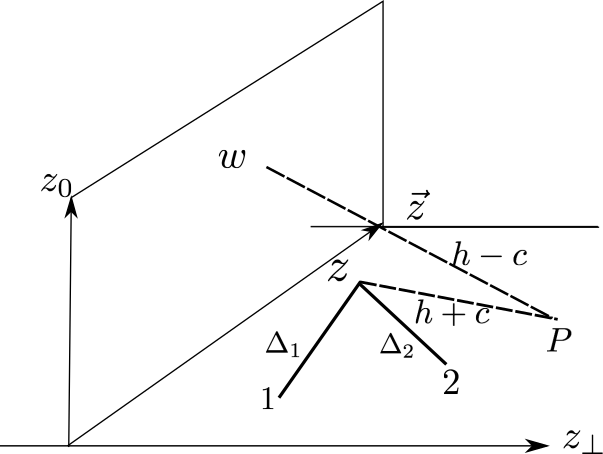}
\caption{Using the split representation of the bulk-to-bulk propagator the bulk exchange Witten diagram is reduced to the product of a three-point contact Witten diagram and an one-point contact Witten diagram.}
\label{fblksplit}
\end{center}
\end{figure}

To begin, we use the spectral representation of the
 bulk-to-bulk propagator \cite{Penedones:2010ue},
\begin{equation}
G_{BB}^\Delta(Z,W)=\int_{-i\infty}^{i\infty} \frac{dc}{(\Delta-h)^2-c^2}\frac{\Gamma(h+c)\Gamma(h-c)}{2\pi^{2h}\Gamma(c)\Gamma(-c)}\int dP (-2P\cdot Z)^{h+c}(-2P\cdot W)^{h-c}\, ,
\end{equation}
where  $h=\frac{d}{2}$. Using this representation, 
 the exchange Witten diagram is written as a product of three-point contact Witten diagram in $AdS_{d+1}$ and an one-point function. There is a common point $P$ sitting on the boundary of $AdS_{d+1}$ which is integrated over. This is schematically represented by Figure \ref{fblksplit}. 

Explicitly, denoting the three-point function by $\langle O_{\Delta_1}(P_1)O_{\Delta_2}(P_2)O_{h+c}(P)\rangle$ and the one-point function by $\langle O_{h-c}(P)\rangle$, the Witten diagram is given by
\begin{equation}
W_{\rm bulk}=\int dP\int dc \frac{\langle O_{\Delta_1}(P_1)O_{\Delta_2}(P_2)O_{h+c}(P)\rangle \langle O_{h-c}(P)\rangle}{(\Delta-h)^2-c^2}\frac{\Gamma(h+c)\Gamma(h-c)}{2\pi^{2h}\Gamma(c)\Gamma(-c)}\;.
\end{equation}
The three-point Witten diagram can be easily evaluated,
\begin{equation}
\begin{split}
\langle O_{\Delta_1}(P_1){}&O_{\Delta_2}(P_2)O_{h+c}(P)\rangle= \frac{\pi^h}{2} \frac{\Gamma(\frac{\Delta_1+\Delta_2+h+c}{2}-2)}{\Gamma(\Delta_1)\Gamma(\Delta_2)\Gamma(h+c)}
 \\
\times{}& \frac{\Gamma(\frac{\Delta_1-\Delta_2+h+c}{2})\Gamma(\frac{\Delta_1+\Delta_2-h-c}{2})\Gamma(\frac{-\Delta_1+\Delta_2+h+c}{2})}{(-P_1\cdot P_2)^{\Delta_2+\Delta_1-h-c}(-P_2\cdot P)^{h+c+\Delta_2-\Delta_1}(-2P_1\cdot P)^{\Delta_1-\Delta_2+h+c}}\,.
\end{split}
\end{equation}
The one-point function is
\begin{equation}
\langle O_{h-c}(P)\rangle=\frac{\pi^{h-\frac{1}{2}}}{2}\frac{\Gamma(\frac{h-c}{2})\Gamma(\frac{-h-c+1)}{2})}{\Gamma(h-c)}\frac{1}{(P\cdot B)^{h-c}}\;.
\end{equation} 
Then the only non-trivial integral remains to be evaluated is the $P$-integral 
\begin{equation}
\begin{split}
{}&\int_{-\infty}^{+\infty}dp_\perp\int_{-\infty}^{+\infty}d\vec{p} \frac{1}{(x_2-p)^{h+c+\Delta_2-\Delta_1}}\frac{1}{(x_1-p)^{h+c+\Delta_1-\Delta_2}}\frac{1}{p_\perp^{h-c}}\\
{}&=\frac{\pi^h}{\Gamma(\frac{h+c+\Delta_1-\Delta_2}{2})\Gamma(\frac{h+c+\Delta_2-\Delta_1}{2})\Gamma(\frac{h-c}{2})} \int \frac{ds}{s}\frac{dt}{t}\frac{du}{u} s^{\frac{h-c}{2}} t^{\frac{h+c+\Delta_1-\Delta_2}{2}} u^{\frac{h+c+\Delta_2-\Delta_1}{2}}\\
{}&\times (u+t)^{-h+\frac{1}{2}}(s+t+u)^{-\frac{1}{2}}\times \exp\left(-\frac{tu(x-y)^2}{t+u}-\frac{s(tx_\perp+uy_\perp)^2}{(t+u)(s+t+u)}\right)\;.
\end{split}
\end{equation} 
We insert into the integral the identity
\begin{equation}
1= \int d\lambda\;\delta(\lambda-(s+t+u))\int d\rho\;\delta(\rho-(t+u))
\end{equation}
and rescale first $t\to \rho t$, $u\to \rho u$, followed by $\lambda\to\lambda \rho$ and the use of an inverse Mellin transformation on the $(tx_\perp+uy_\perp)^2$ exponent. The integrals then become elementary. The final result for the bulk exchange Witten diagram is
\begin{equation}\label{blkmellinspec}
\begin{split}
W_{\rm bulk}={}&\frac{\pi^{h-\frac{1}{2}}}{2 \Gamma(\Delta_1)\Gamma(\Delta_2)x_\perp^{\Delta_1}y_\perp^{\Delta_2}}\int_{-i\infty}^{+i\infty} dc \int_{-i\infty}^{+i\infty} d\tau \frac{f(c,\tau)f(-c,\tau)}{(\Delta-h)^2-c^2}\eta^{\tau-\frac{\Delta_1+\Delta_2}{2}}\\
{}&\times \frac{\Gamma(\tau)\Gamma(\tau+\frac{\Delta_1-\Delta_2}{2})\Gamma(\tau+\frac{\Delta_2-\Delta_1}{2})}{\Gamma(\frac{1}{2}-\tau)\Gamma(2\tau)} \, ,
\end{split}
\end{equation}
where $h=d/2$ and
\begin{equation}
f(c,\tau)=\frac{\Gamma(\frac{\Delta_1+\Delta_2-h+c}{2})\Gamma(\frac{1+c-h}{2})\Gamma(\frac{h+c}{2}-\tau)}{2\Gamma(c)}\;.
\end{equation}

It is not difficult to recognize in the above expression poles expected from the operator product expansion. Closing the $c$-contour to the right on the complex plane, the following poles in $c$ are encircled:
\begin{enumerate}
\item[I.] $c=\Delta-h$.
\item[II.] $c=h-2\tau+2m_{II}$ with integer $m_{II}\geq 0$.
\item[III.] $c=1-h+2m_{III}$ with integer $m_{III}\geq 0$.
\item[IV.] $c=-h+\Delta_1+\Delta_2+2m_{IV}$ with integer $m_{IV}\geq 0$.
\end{enumerate}
Taking residues at these $c$-poles, the poles I and II give rise to a series of simple poles at $\tau=\frac{\Delta}{2}+n_{\rm st}$ with integer $n_{\rm st}\geq 0$. This family of simple poles corresponds to the exchanged single-trace operator $O_\Delta$, as can be seen from (\ref{bulkcb}). Poles in $c$ from II and IV lead to another series of poles at $\tau=\frac{\Delta_1+\Delta_2}{2}+n_{\rm dt}$ with integer $n_{\rm dt}\geq 0$ and they amount to the exchanged double-trace operators from the bulk-OPE $:O_{\Delta_1}\square^{n_{\rm dt}}O_{\Delta_2}:$. There are further $\tau$-poles from the factor $\Gamma(\tau+\frac{\Delta_1-\Delta_2}{2})\Gamma(\tau+\frac{\Delta_2-\Delta_1}{2})$  in (\ref{blkmellinspec}) which are independent of the $c$-poles. These two families of poles at $\tau=\pm\frac{\Delta_1-\Delta_2}{2}+n_{\rm int}$ with integer $n_{\rm int}\geq 0$ lie in the opposite direction on $\tau$-plane compared to the aforementioned series. According to (\ref{boundarycb}) such poles correspond to the operator exchange of $\partial_{\perp}^{n_{\rm int}}O_{\Delta_{1,2}}(x_{1,2\perp}=0)$ in the interface channel.

\subsection{Interface exchange Witten diagrams: the truncation method}\label{bdrtrunc} 
In this subsection we evaluate the interface exchange Witten diagram using a method analogous to the one used in Section \ref{blktrunc}.

The interface exchange Witten diagram represented by Figure \ref{fboundaryx} is  given by the following integral,
\begin{equation}
W_{\rm interface}=\int_{AdS_d}dW_1\int_{AdS_d}dW_2\; G_{B\partial}^{\Delta_1}(P_1,W_1)\;G_{BB,\; AdS_d}^{\Delta}(W_1,W_2)\;G_{B\partial}^{\Delta_2}(P_2,W_2)\;.
\end{equation}
We focus on the integral of $W_1$ denoted as
\begin{equation}
A(P_1,W_2)=\int_{AdS_d}dW_1\;G_{B\partial}^{\Delta_1}(P_1,W_1)\;G_{BB,\; AdS_d}^{\Delta}(W_1,W_2)\;.
\end{equation}
This integral has $AdS_{d}$ isometry and should depend on a single variable $t$ invariant under the scaling $w_2\to \lambda w_2$, $x_1\to \lambda x_1$
\begin{equation}\label{tboundary}
t\equiv \frac{P_1\cdot W_2}{P_1\cdot B}=\frac{w_{2,0}^2+x_{1,\perp}^2+(\vec{w}_2-\vec{x}_1)^2}{w_{2,0} x_{1,\perp}}\;.
\end{equation}
The function $A(P_1,W_2)$ therefore takes the form
\begin{equation}
A(P_1,W_2)=x_{1,\perp}^{-\Delta_1}f(t)\;.
\end{equation}
To work out $f(t)$, we use the equation of motion for the bulk-to-bulk propagator inside $AdS_d$. It leads to the following equation
\begin{equation}
(-\square_{W_2,AdS_d}+m^2)(x_{1,\perp}^{-\Delta_1}f(t))=x_{1,\perp}^{-\Delta_1}t^{-\Delta_1}
\end{equation}
where $m^2=\Delta(\Delta-(d-1))$. The Laplacian acts on a function of $t$ as
\begin{equation}
\square_{W_2,AdS_d} f(t)=(t^2-4)f''(t)+d tf'(t)\;.
\end{equation}
The function $f(t)$ also admits a polynomial solution when $\Delta<\Delta_1$ and has even integer difference,
\begin{equation}
f(t)=\sum_{k_{\rm min}}^{k_{\rm max}}a_k t^k \, ,
\end{equation}
where
\begin{equation}
\begin{split}
{}& a_{k+2}=\frac{(k+\Delta)(k-(\Delta-(d-1)))}{4(k+1)(k+2)}a_k\;,\\
{}& {k_{\rm min}}=-\Delta_1+2\;,\\
{}& {k_{\rm max}}= -\Delta\;,\\
{}& a_{k_{\rm min}}=\frac{1}{4(-\Delta_1+2)(-\Delta_1+1)}\;.
\end{split}
\end{equation}

\begin{figure}[t]
\begin{center}
\includegraphics[scale=0.5]{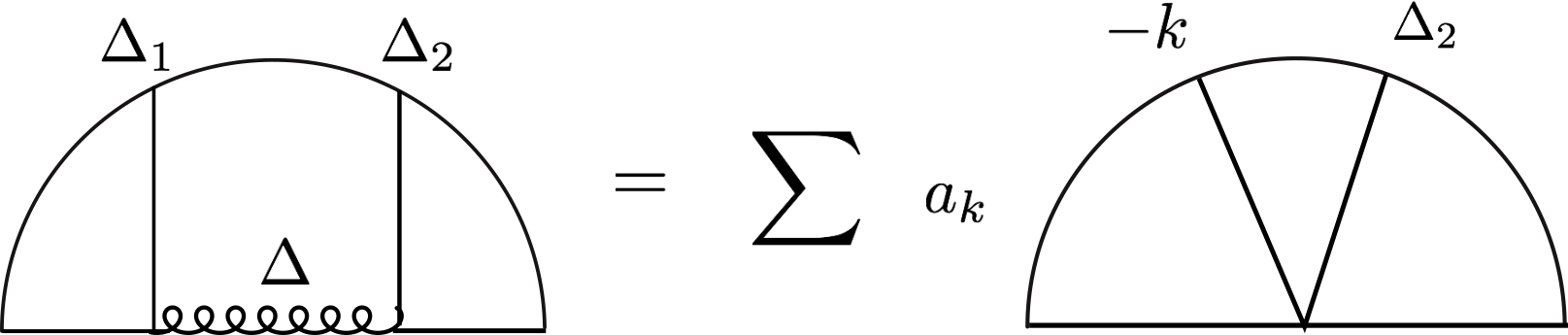}
\caption{The interface exchange Witten diagram is replaced by a finite sum of contact Witten diagrams when $\Delta_1-\Delta$ is a positive even integer.}
\label{fbdrtrunc}
\end{center}
\end{figure}

Using the definition of $t$ in (\ref{tboundary}), we find that each monomial of $t$ corresponds to a contact vertex. The polynomial solution to $f(t)$ means we can express the exchange Witten diagram as a sum of contact Witten diagrams (Figure \ref{fbdrtrunc}),
\begin{equation}
W_{\rm interface}=\int_{AdS_d}dW_2\sum_{k_{\rm min}}^{k_{\rm max}}a_k x_{1,\perp}^{-\Delta_1-k}\; G_{B\partial}^{-k}(P_1,W_2)\;G_{B\partial}^{\Delta_2}(P_2,W_2)\;.
\end{equation} 
Using the Mellin formula for contact Witten diagrams (\ref{2ptcontact}), we get the Mellin amplitude for the interface exchange Witten diagram
\begin{equation}\label{bdrmellintruc}
\mathcal{M}_{\rm interface}=\pi^{\frac{d}{2}}\sum_{k_{\rm min}}^{k_{\rm max}} a_k 2^{k+\Delta_1} \frac{\Gamma(\frac{\Delta_2-k-(d-1)}{2})}{\Gamma(-k)\Gamma(\Delta_2))}\times \frac{\Gamma(-k-\tau)\Gamma(\frac{\Delta_1+\Delta_2+1}{2}-\tau)}{\Gamma(\Delta_1-\tau)\Gamma(\frac{-k+\Delta_2+1}{2}-\tau)}\;.
\end{equation}

\subsection{Interface exchange Witten diagrams: the spectral representation}\label{bdrspec} 
\begin{figure}[t]
\begin{center}
\includegraphics[scale=0.6]{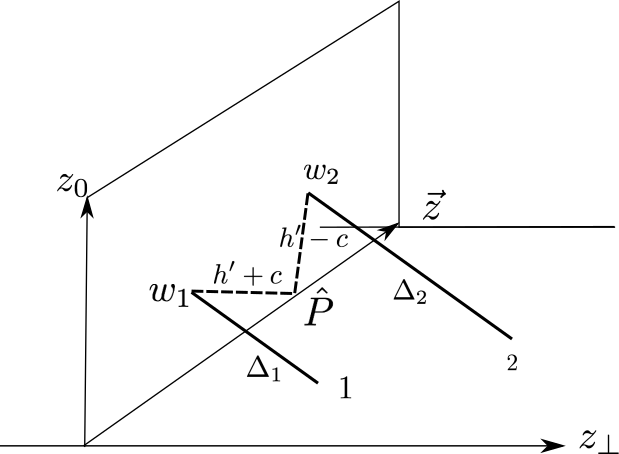}
\caption{Using the split representation of the bulk-to-bulk propagator the interface exchange Witten diagram is reduced to the product of two bulk-interface two-point contact Witten diagrams.}
\label{fbdrsplit}
\end{center}
\end{figure}

In this section we compute the interface exchange Witten diagram using the spectral representation. We use the split form of the $AdS_d$ propagator  
\begin{equation}
G_{BB,\; AdS_d}^\Delta=\int_{i\infty}^{i\infty} \frac{dc}{(\Delta-h')^2-c^2}\frac{\Gamma(h'+c)\Gamma(h'-c)}{2\pi^{2h'}\Gamma(c)\Gamma(-c)}\int d\widehat{P} (-2\widehat{P}\cdot W_1)^{h'+c}(-2\widehat{P}\cdot W_2)^{h'-c}
\end{equation}
with $h'=(d-1)/2$ to write the Witten diagram into the product of two bulk-interface two-point functions $W_{1,1}$ that we studied in Section \ref{blkbdr2pt}. This splitting is schematically illustrated in Figure \ref{fbdrsplit}. Notice the point $\widehat{P}$ being integrated over is sitting at the boundary of $AdS_d$. Denoting the two-point functions by $\langle O_{\Delta_1}(P_1) \widehat{O}_{h'+c}(\widehat{P}) \rangle$ and $\langle O_{\Delta_2}(P_2) \widehat{O}_{h'-c}(\widehat{P}) \rangle$, the Witten diagram now takes the form
\begin{equation}
W_{\rm interface}=\int d\widehat{P}\int dc \frac{\langle O_{\Delta_1}(P_1) \widehat{O}_{h'+c}(\widehat{P}) \langle O_{\Delta_2}(P_2) \widehat{O}_{h'-c}(\widehat{P}) \rangle}{(\Delta-h')^2-c^2}\frac{\Gamma(h'+c)\Gamma(h'-c)}{2\pi^{2h'}\Gamma(c)\Gamma(-c)}\;.
\end{equation}
The two-point functions have been worked out and are given by (\ref{W11}). The only integral we need to do is the $\widehat{P}$-integral
\begin{equation}
\int d^{d-1}\vec{p}\;(x_{1,\perp}^2+(\vec{x}_1-\vec{p})^2)^{-(h'+c)}(x_{2,\perp}^2+(\vec{x}_2-\vec{p})^2)^{-(h'-c)}\;.
\end{equation}
Evaluating this integral presents little difficulty using the techniques we have developed in the previous sections. The answer is simply
\begin{equation}
\frac{\pi^{h'}}{\Gamma(h'+c)\Gamma(h'-c)}\int_{-i\infty}^{i\infty}d\tau \eta^{-\tau}\frac{\Gamma(\tau)\Gamma(h'-\tau)\Gamma(h'+c-\tau)\Gamma(h'-c-\tau)}{(x_{1,\perp})^{h'+c}(x_{2,\perp})^{h'-c} \Gamma(2h'-2\tau)}\;.
\end{equation}
Hence the interface exchange Witten diagram is given by the following spectral representation
 \begin{equation}\label{bdrmellinspec}
\begin{split}
W_{\rm interface}={}&\frac{\pi^{h'}}{2\Gamma(\Delta_1)\Gamma(\Delta_2)x_\perp^{\Delta_1}y_\perp^{\Delta_2}}\int_{-i\infty}^{+i\infty} d\tau \eta^{-\tau}\frac{\Gamma(\tau)\Gamma(h'-\tau)}{\Gamma(2h'-2\tau)}\\
{}&\times \int_{-i\infty}^{+i\infty} dc \frac{1}{(\Delta-h')^2-c^2}f(c,\tau)f(-c,\tau)
\end{split}
\end{equation}
where $h'=(d-1)/2$ and
\begin{equation}
f(c,\tau)=\frac{\Gamma(h'+c-\tau)\Gamma(\frac{\Delta_1-h'+c}{2})\Gamma(\frac{\Delta_2-h'+c}{2})}{2\Gamma(c)}\;.
\end{equation}

The pole structure in the above expression is again consistent with the OPE expectation. Closing the $c$-contour to the right, we encircle the following $c$-poles:
\begin{enumerate}
\item[I.] $c=\Delta-h'$.
\item[II.] $c=h'-\tau+m_{II}$ with integer $m_{II}\geq 0$.
\item[III.] $c=\Delta_1-h'+2m_{III}$ with integer $m_{III}\geq 0$.
\item[IV.] $c=\Delta_2-h'+2m_{IV}$ with integer $m_{IV}\geq 0$.\end{enumerate}
The residues from poles I and II lead to a series of simple poles at $\tau=\Delta+n_{\rm st}$ with integer $n_{\rm st}\geq 0$ and they are identified with the interface exchange of operator $\widehat{O}_{\Delta}$ by (\ref{boundarycb}). The residues at poles III and IV result in two series simple poles at $\tau=\Delta_{1,2}+n_{\rm int}$ where the integer $n_{\rm int}\geq 0$. These poles correspond to the exchange of interface operators $\partial_\perp^{n_{\rm int}}O_{\Delta_{1,2}}(x_{1,2\perp}=0)$. Finally, there is a family of poles at $\tau=-n_{\rm dt}$ from the factor $\Gamma(\tau)$ in (\ref{bdrmellinspec}) with integer $n_{\rm dt}\geq0$. By comparing with (\ref{bulkcb}), we find they correspond to the bulk exchange of double-trace operators $:O_{\Delta_1}\square^{n_{\rm dt}}O_{\Delta_2}:$.

\section{Conclusions}

\label{conclusions}

We have extended the Mellin representation of CFT$_d$ correlators to allow for boundaries and interfaces, and demonstrated
its power in a  simple holographic setup. This generalization works very naturally. The Mandelstam-like variables that appear in the Mellin amplitude can be interpreted as 
the kinematic invariants of a scattering process off a fixed codimension-one target in $d+1$ dimensional spacetime, as one expects from the holographic intuition. 
For theories that do admit an actual holographic dual in $AdS_{d+1}$, the Mellin representation leads to great simplifications in the evaluation of Witten diagrams.
With an appropriate definition of the Mellin transform, contact diagrams are just constant in Mellin space, while exchange diagrams are meromorphic functions with simple poles.

In this paper we have laid out the technical groundwork necessary for applications of the Mellin formalism to several physical questions  in boundary and interface CFT.
In an ongoing project  \cite{inprogress1}, we are extending the ``Mellin-bootstrap''  approach of \cite{Gopakumar:2016cpb,Gopakumar:2016wkt,Dey:2016mcs}  to derive the $\varepsilon$-expansion of BCFTs
that can be described by a Wilson-Fisher fixed point.  The conceptual underpinnings of this approach ({\it e.g.}, the
question of {\it why} exchange Witten diagrams constitute a good basis) may be easier to study in this simpler setting. Another direction that we are pursuing is the generalization of
the philosophy of  \cite{Rastelli:2016nze}  to the calculation of holographic two-point functions of one-half BPS operators in the holographic setting of \cite{DeWolfe:2001pq}. We are hopeful that the constraints of superconformal symmetry and analyticity will determine uniquely the form of these correlators. 

We have focussed here on correlators of scalar operators. The Mellin representation for spinning correlators has not been fully developed even in the standard case of a CFT with no defects,
see \cite{Paulos:2011ie,Goncalves:2014rfa} for the state of the art. 
In fact, the simplest spinning correlators in ICFT (bulk two-point functions)  should be  more  tractable than the  simplest spinning correlators in the standard case (four-point functions), and  serve as
a convenient beachhead to attack the general problem. Finally, it would be natural and interesting to extend the Mellin formalism to CFTs endowed with  conformal defects of arbitrary codimension. 

\acknowledgments
Our work is supported in part by NSF Grant PHY-1620628.

\bibliography{bcft} 

\providecommand{\href}[2]{#2}\begingroup\raggedright\begin{thebibliography}{10}

\bibitem{Mack:2009mi}
G.~Mack, ``{D-independent representation of Conformal Field Theories in D
  dimensions via transformation to auxiliary Dual Resonance Models. Scalar
  amplitudes},''
\href{http://arxiv.org/abs/0907.2407}{{\ttfamily arXiv:0907.2407 [hep-th]}}.

\bibitem{Penedones:2010ue}
J.~Penedones, ``{Writing CFT correlation functions as AdS scattering
  amplitudes},'' \href{http://dx.doi.org/10.1007/JHEP03(2011)025}{{\em JHEP}
  {\bfseries 03} (2011) 025},
\href{http://arxiv.org/abs/1011.1485}{{\ttfamily arXiv:1011.1485 [hep-th]}}.

\bibitem{Paulos:2016fap}
M.~F. Paulos, J.~Penedones, J.~Toledo, B.~C. van Rees, and P.~Vieira, ``{The
  S-matrix Bootstrap I: QFT in AdS},''
\href{http://arxiv.org/abs/1607.06109}{{\ttfamily arXiv:1607.06109 [hep-th]}}.

\bibitem{Paulos:2011ie}
M.~F. Paulos, ``{Towards Feynman rules for Mellin amplitudes},''
  \href{http://dx.doi.org/10.1007/JHEP10(2011)074}{{\em JHEP} {\bfseries 10}
  (2011) 074},
\href{http://arxiv.org/abs/1107.1504}{{\ttfamily arXiv:1107.1504 [hep-th]}}.

\bibitem{Fitzpatrick:2011ia}
A.~L. Fitzpatrick, J.~Kaplan, J.~Penedones, S.~Raju, and B.~C. van Rees, ``{A
  Natural Language for AdS/CFT Correlators},''
  \href{http://dx.doi.org/10.1007/JHEP11(2011)095}{{\em JHEP} {\bfseries 11}
  (2011) 095},
\href{http://arxiv.org/abs/1107.1499}{{\ttfamily arXiv:1107.1499 [hep-th]}}.

\bibitem{Costa:2014kfa}
M.~S. Costa, V.~Gon{\c c}alves, and J.~Penedones, ``{Spinning AdS
  Propagators},'' \href{http://dx.doi.org/10.1007/JHEP09(2014)064}{{\em JHEP}
  {\bfseries 09} (2014) 064},
\href{http://arxiv.org/abs/1404.5625}{{\ttfamily arXiv:1404.5625 [hep-th]}}.

\bibitem{Goncalves:2014rfa}
V.~Gon{\c c}alves, J.~Penedones, and E.~Trevisani, ``{Factorization of Mellin
  amplitudes},'' \href{http://dx.doi.org/10.1007/JHEP10(2015)040}{{\em JHEP}
  {\bfseries 10} (2015) 040},
\href{http://arxiv.org/abs/1410.4185}{{\ttfamily arXiv:1410.4185 [hep-th]}}.

\bibitem{Aharony:2016dwx}
O.~Aharony, L.~F. Alday, A.~Bissi, and E.~Perlmutter, ``{Loops in AdS from
  Conformal Field Theory},''
\href{http://arxiv.org/abs/1612.03891}{{\ttfamily arXiv:1612.03891 [hep-th]}}.

\bibitem{Rastelli:2016nze}
L.~Rastelli and X.~Zhou, ``{Mellin amplitudes for $AdS_5\times S^5$},''
  \href{http://dx.doi.org/10.1103/PhysRevLett.118.091602}{{\em Phys. Rev.
  Lett.} {\bfseries 118} no.~9, (2017) 091602},
\href{http://arxiv.org/abs/1608.06624}{{\ttfamily arXiv:1608.06624 [hep-th]}}.

\bibitem{Paulos:2012nu}
M.~F. Paulos, M.~Spradlin, and A.~Volovich, ``{Mellin Amplitudes for Dual
  Conformal Integrals},'' \href{http://dx.doi.org/10.1007/JHEP08(2012)072}{{\em
  JHEP} {\bfseries 08} (2012) 072},
\href{http://arxiv.org/abs/1203.6362}{{\ttfamily arXiv:1203.6362 [hep-th]}}.

\bibitem{Fitzpatrick:2012cg}
A.~L. Fitzpatrick and J.~Kaplan, ``{AdS Field Theory from Conformal Field
  Theory},'' \href{http://dx.doi.org/10.1007/JHEP02(2013)054}{{\em JHEP}
  {\bfseries 02} (2013) 054},
\href{http://arxiv.org/abs/1208.0337}{{\ttfamily arXiv:1208.0337 [hep-th]}}.

\bibitem{Costa:2012cb}
M.~S. Costa, V.~Goncalves, and J.~Penedones, ``{Conformal Regge theory},''
  \href{http://dx.doi.org/10.1007/JHEP12(2012)091}{{\em JHEP} {\bfseries 12}
  (2012) 091},
\href{http://arxiv.org/abs/1209.4355}{{\ttfamily arXiv:1209.4355 [hep-th]}}.

\bibitem{Alday:2014tsa}
L.~F. Alday, A.~Bissi, and T.~Lukowski, ``{Lessons from crossing symmetry at
  large N},'' \href{http://dx.doi.org/10.1007/JHEP06(2015)074}{{\em JHEP}
  {\bfseries 06} (2015) 074},
\href{http://arxiv.org/abs/1410.4717}{{\ttfamily arXiv:1410.4717 [hep-th]}}.

\bibitem{Goncalves:2014ffa}
V.~Gon{\c c}alves, ``{Four point function of $\mathcal{N}=4$ stress-tensor
  multiplet at strong coupling},''
  \href{http://dx.doi.org/10.1007/JHEP04(2015)150}{{\em JHEP} {\bfseries 04}
  (2015) 150},
\href{http://arxiv.org/abs/1411.1675}{{\ttfamily arXiv:1411.1675 [hep-th]}}.

\bibitem{Alday:2016htq}
L.~F. Alday and A.~Bissi, ``{Unitarity and positivity constraints for CFT at
  large central charge},''
\href{http://arxiv.org/abs/1606.09593}{{\ttfamily arXiv:1606.09593 [hep-th]}}.

\bibitem{Gopakumar:2016wkt}
R.~Gopakumar, A.~Kaviraj, K.~Sen, and A.~Sinha, ``{Conformal Bootstrap in
  Mellin Space},'' \href{http://dx.doi.org/10.1103/PhysRevLett.118.081601}{{\em
  Phys. Rev. Lett.} {\bfseries 118} no.~8, (2017) 081601},
\href{http://arxiv.org/abs/1609.00572}{{\ttfamily arXiv:1609.00572 [hep-th]}}.

\bibitem{Gopakumar:2016cpb}
R.~Gopakumar, A.~Kaviraj, K.~Sen, and A.~Sinha, ``{A Mellin space approach to
  the conformal bootstrap},''
\href{http://arxiv.org/abs/1611.08407}{{\ttfamily arXiv:1611.08407 [hep-th]}}.

\bibitem{Dey:2016mcs}
P.~Dey, A.~Kaviraj, and A.~Sinha, ``{Mellin space bootstrap for global
  symmetry},''
\href{http://arxiv.org/abs/1612.05032}{{\ttfamily arXiv:1612.05032 [hep-th]}}.

\bibitem{Cardy:1989ir}
J.~L. Cardy, ``{Boundary Conditions, Fusion Rules and the Verlinde Formula},''
\href{http://dx.doi.org/10.1016/0550-3213(89)90521-X}{{\em Nucl. Phys.}
  {\bfseries B324} (1989) 581--596}.

\bibitem{Cardy:1991tv}
J.~L. Cardy and D.~C. Lewellen, ``{Bulk and boundary operators in conformal
  field theory},''
\href{http://dx.doi.org/10.1016/0370-2693(91)90828-E}{{\em Phys. Lett.}
  {\bfseries B259} (1991) 274--278}.

\bibitem{Bachas:2001vj}
C.~Bachas, J.~de~Boer, R.~Dijkgraaf, and H.~Ooguri, ``{Permeable conformal
  walls and holography},''
  \href{http://dx.doi.org/10.1088/1126-6708/2002/06/027}{{\em JHEP} {\bfseries
  06} (2002) 027},
\href{http://arxiv.org/abs/hep-th/0111210}{{\ttfamily arXiv:hep-th/0111210
  [hep-th]}}.

\bibitem{Cardy:1984bb}
J.~L. Cardy, ``{Conformal Invariance and Surface Critical Behavior},''
\href{http://dx.doi.org/10.1016/0550-3213(84)90241-4}{{\em Nucl. Phys.}
  {\bfseries B240} (1984) 514--532}.

\bibitem{Gaiotto:2008sa}
D.~Gaiotto and E.~Witten, ``{Supersymmetric Boundary Conditions in N=4 Super
  Yang-Mills Theory},'' \href{http://dx.doi.org/10.1007/s10955-009-9687-3}{{\em
  J. Statist. Phys.} {\bfseries 135} (2009) 789--855},
\href{http://arxiv.org/abs/0804.2902}{{\ttfamily arXiv:0804.2902 [hep-th]}}.

\bibitem{Karch:2000gx}
A.~Karch and L.~Randall, ``{Open and closed string interpretation of SUSY CFT's
  on branes with boundaries},''
  \href{http://dx.doi.org/10.1088/1126-6708/2001/06/063}{{\em JHEP} {\bfseries
  06} (2001) 063},
\href{http://arxiv.org/abs/hep-th/0105132}{{\ttfamily arXiv:hep-th/0105132
  [hep-th]}}.

\bibitem{DeWolfe:2001pq}
O.~DeWolfe, D.~Z. Freedman, and H.~Ooguri, ``{Holography and defect conformal
  field theories},'' \href{http://dx.doi.org/10.1103/PhysRevD.66.025009}{{\em
  Phys. Rev.} {\bfseries D66} (2002) 025009},
\href{http://arxiv.org/abs/hep-th/0111135}{{\ttfamily arXiv:hep-th/0111135
  [hep-th]}}.

\bibitem{Rattazzi:2008pe}
R.~Rattazzi, V.~S. Rychkov, E.~Tonni, and A.~Vichi, ``{Bounding scalar operator
  dimensions in 4D CFT},''
  \href{http://dx.doi.org/10.1088/1126-6708/2008/12/031}{{\em JHEP} {\bfseries
  12} (2008) 031},
\href{http://arxiv.org/abs/0807.0004}{{\ttfamily arXiv:0807.0004 [hep-th]}}.

\bibitem{Rychkov:2016iqz}
S.~Rychkov, \href{http://dx.doi.org/10.1007/978-3-319-43626-5}{{\em {EPFL
  Lectures on Conformal Field Theory in D>= 3 Dimensions}}}.
\newblock SpringerBriefs in Physics. 2016.
\newblock
\href{http://arxiv.org/abs/1601.05000}{{\ttfamily arXiv:1601.05000 [hep-th]}}.
\newblock

\bibitem{Simmons-Duffin:2016gjk}
D.~Simmons-Duffin, \href{http://dx.doi.org/10.1142/9789813149441_0001}{``{The
  Conformal Bootstrap},''} in {\em {Proceedings, Theoretical Advanced Study
  Institute in Elementary Particle Physics: New Frontiers in Fields and Strings
  (TASI 2015): Boulder, CO, USA, June 1-26, 2015}}, pp.~1--74.
\newblock 2017.
\newblock
\href{http://arxiv.org/abs/1602.07982}{{\ttfamily arXiv:1602.07982 [hep-th]}}.
\newblock

\bibitem{Liendo:2012hy}
P.~Liendo, L.~Rastelli, and B.~C. van Rees, ``{The Bootstrap Program for
  Boundary $CFT_d$},'' \href{http://dx.doi.org/10.1007/JHEP07(2013)113}{{\em
  JHEP} {\bfseries 07} (2013) 113},
\href{http://arxiv.org/abs/1210.4258}{{\ttfamily arXiv:1210.4258 [hep-th]}}.

\bibitem{Gliozzi:2013ysa}
F.~Gliozzi, ``{More constraining conformal bootstrap},''
  \href{http://dx.doi.org/10.1103/PhysRevLett.111.161602}{{\em Phys. Rev.
  Lett.} {\bfseries 111} (2013) 161602},
\href{http://arxiv.org/abs/1307.3111}{{\ttfamily arXiv:1307.3111 [hep-th]}}.

\bibitem{Gliozzi:2015qsa}
F.~Gliozzi, P.~Liendo, M.~Meineri, and A.~Rago, ``{Boundary and Interface CFTs
  from the Conformal Bootstrap},''
  \href{http://dx.doi.org/10.1007/JHEP05(2015)036}{{\em JHEP} {\bfseries 05}
  (2015) 036},
\href{http://arxiv.org/abs/1502.07217}{{\ttfamily arXiv:1502.07217 [hep-th]}}.

\bibitem{Billo:2016cpy}
M.~Bill{\`o}, V.~Gon{\c c}alves, E.~Lauria, and M.~Meineri, ``{Defects in
  conformal field theory},''
  \href{http://dx.doi.org/10.1007/JHEP04(2016)091}{{\em JHEP} {\bfseries 04}
  (2016) 091},
\href{http://arxiv.org/abs/1601.02883}{{\ttfamily arXiv:1601.02883 [hep-th]}}.

\bibitem{Gliozzi:2016cmg}
F.~Gliozzi, ``{Truncatable bootstrap equations in algebraic form and critical
  surface exponents},'' \href{http://dx.doi.org/10.1007/JHEP10(2016)037}{{\em
  JHEP} {\bfseries 10} (2016) 037},
\href{http://arxiv.org/abs/1605.04175}{{\ttfamily arXiv:1605.04175 [hep-th]}}.

\bibitem{Liendo:2016ymz}
P.~Liendo and C.~Meneghelli, ``{Bootstrap equations for $ \mathcal{N} $ = 4 SYM
  with defects},'' \href{http://dx.doi.org/10.1007/JHEP01(2017)122}{{\em JHEP}
  {\bfseries 01} (2017) 122},
\href{http://arxiv.org/abs/1608.05126}{{\ttfamily arXiv:1608.05126 [hep-th]}}.

\bibitem{Karch:2001cw}
A.~Karch and L.~Randall, ``{Localized gravity in string theory},''
  \href{http://dx.doi.org/10.1103/PhysRevLett.87.061601}{{\em Phys. Rev. Lett.}
  {\bfseries 87} (2001) 061601},
\href{http://arxiv.org/abs/hep-th/0105108}{{\ttfamily arXiv:hep-th/0105108
  [hep-th]}}.

\bibitem{Hijano:2015zsa}
E.~Hijano, P.~Kraus, E.~Perlmutter, and R.~Snively, ``{Witten Diagrams
  Revisited: The AdS Geometry of Conformal Blocks},''
  \href{http://dx.doi.org/10.1007/JHEP01(2016)146}{{\em JHEP} {\bfseries 01}
  (2016) 146},
\href{http://arxiv.org/abs/1508.00501}{{\ttfamily arXiv:1508.00501 [hep-th]}}.

\bibitem{Aharony:2003qf}
O.~Aharony, O.~DeWolfe, D.~Z. Freedman, and A.~Karch, ``{Defect conformal field
  theory and locally localized gravity},''
  \href{http://dx.doi.org/10.1088/1126-6708/2003/07/030}{{\em JHEP} {\bfseries
  07} (2003) 030},
\href{http://arxiv.org/abs/hep-th/0303249}{{\ttfamily arXiv:hep-th/0303249
  [hep-th]}}.

\bibitem{Fitzpatrick:2012yx}
A.~L. Fitzpatrick, J.~Kaplan, D.~Poland, and D.~Simmons-Duffin, ``{The Analytic
  Bootstrap and AdS Superhorizon Locality},''
  \href{http://dx.doi.org/10.1007/JHEP12(2013)004}{{\em JHEP} {\bfseries 12}
  (2013) 004},
\href{http://arxiv.org/abs/1212.3616}{{\ttfamily arXiv:1212.3616 [hep-th]}}.

\bibitem{Komargodski:2012ek}
Z.~Komargodski and A.~Zhiboedov, ``{Convexity and Liberation at Large Spin},''
  \href{http://dx.doi.org/10.1007/JHEP11(2013)140}{{\em JHEP} {\bfseries 11}
  (2013) 140},
\href{http://arxiv.org/abs/1212.4103}{{\ttfamily arXiv:1212.4103 [hep-th]}}.

\bibitem{longpaper}
L.~Rastelli and X.~Zhou, ``{How to Succeed at Holographic Correlators Without
  Really Trying},''
\href{http://arxiv.org/abs/1710.05923}{{\ttfamily arXiv:1710.05923 [hep-th]}}.

\bibitem{Witten:1998qj}
E.~Witten, ``{Anti-de Sitter space and holography},'' {\em Adv. Theor. Math.
  Phys.} {\bfseries 2} (1998) 253--291,
\href{http://arxiv.org/abs/hep-th/9802150}{{\ttfamily arXiv:hep-th/9802150
  [hep-th]}}.

\bibitem{Freedman:1998tz}
D.~Z. Freedman, S.~D. Mathur, A.~Matusis, and L.~Rastelli, ``{Correlation
  functions in the CFT(d) / AdS(d+1) correspondence},''
  \href{http://dx.doi.org/10.1016/S0550-3213(99)00053-X}{{\em Nucl. Phys.}
  {\bfseries B546} (1999) 96--118},
\href{http://arxiv.org/abs/hep-th/9804058}{{\ttfamily arXiv:hep-th/9804058
  [hep-th]}}.

\bibitem{Inami:1985wu}
T.~Inami and H.~Ooguri, ``{One Loop Effective Potential in Anti-de Sitter
  Space},''
\href{http://dx.doi.org/10.1143/PTP.73.1051}{{\em Prog. Theor. Phys.}
  {\bfseries 73} (1985) 1051}.

\bibitem{Fronsdal:1974ew}
C.~Fronsdal, ``{Elementary particles in a curved space. ii},''
\href{http://dx.doi.org/10.1103/PhysRevD.10.589}{{\em Phys. Rev.} {\bfseries
  D10} (1974) 589--598}.

\bibitem{Burgess:1985zz}
C.~P. Burgess, A.~Font, and F.~Quevedo, ``{Low-Energy Effective Action for the
  Superstring},''
\href{http://dx.doi.org/10.1016/0550-3213(86)90239-7}{{\em Nucl. Phys.}
  {\bfseries B272} (1986) 661--676}.

\bibitem{Burges:1985qq}
C.~J.~C. Burges, D.~Z. Freedman, S.~Davis, and G.~W. Gibbons, ``{Supersymmetry
  in Anti-de Sitter Space},''
\href{http://dx.doi.org/10.1016/0003-4916(86)90203-4}{{\em Annals Phys.}
  {\bfseries 167} (1986) 285}.

\bibitem{McAvity:1995zd}
D.~M. McAvity and H.~Osborn, ``{Conformal field theories near a boundary in
  general dimensions},''
  \href{http://dx.doi.org/10.1016/0550-3213(95)00476-9}{{\em Nucl. Phys.}
  {\bfseries B455} (1995) 522--576},
\href{http://arxiv.org/abs/cond-mat/9505127}{{\ttfamily arXiv:cond-mat/9505127
  [cond-mat]}}.

\bibitem{DHoker:1999mqo}
E.~D'Hoker, D.~Z. Freedman, and L.~Rastelli, ``{AdS / CFT four point functions:
  How to succeed at z integrals without really trying},''
  \href{http://dx.doi.org/10.1016/S0550-3213(99)00526-X}{{\em Nucl. Phys.}
  {\bfseries B562} (1999) 395--411},
\href{http://arxiv.org/abs/hep-th/9905049}{{\ttfamily arXiv:hep-th/9905049
  [hep-th]}}.

\bibitem{inprogress1}
W.~Peelaers, L.~Rastelli, and X.~Zhou {\em work in progress} .

\end{thebibliography}\endgroup
\bibliographystyle{utphys}

\end{document}